\documentclass[onecolumn,showpacs,preprintnumbers,amsmath,amssymb,nofootinbib,APS]{revtex4}

\usepackage{dcolumn}
\usepackage{bm}
\usepackage[latin1]{inputenc}
\usepackage[spanish,english]{babel}
\usepackage{amsfonts}
\usepackage{amssymb}
\usepackage{graphicx}

\newcommand{\be}{\begin{equation}}
\newcommand{\ee}{\end{equation}}
\newcommand{\bea}{\begin{eqnarray}}
\newcommand{\eea}{\end{eqnarray}}

\begin{document}

\preprint{hep-th/0611355}

\title{{\bf Short-distance contribution to the spectrum of Hawking
radiation}}

\author{I. Agulló$^{a,b}$ and J. Navarro-Salas$^a$}\email{ivan.agullo@uv.es  ,  jnavarro@ific.uv.es}
\affiliation{ {\footnotesize a) Departamento de Física
Teórica and IFIC, Centro Mixto Universidad de Valencia-CSIC.\\
    Facultad de Física, Universidad de Valencia,
        Burjassot-46100, Valencia, Spain.\\ b) Enrico Fermi Institute and Department of Physics,
University of Chicago, Chicago, IL 60637 USA }}

\author{Gonzalo J. Olmo and Leonard Parker}\email{olmoalba@uwm.edu  ,  leonard@uwm.edu}
\affiliation{ {\footnotesize Physics Department, University of Wisconsin-Milwaukee, P.O.Box 413, Milwaukee, WI 53201 USA}}

\date{March 1st, 2007}

\begin{abstract}
The Hawking effect  can be rederived in terms of two-point
functions and in such a way that it makes it possible to estimate,
within the conventional semiclassical theory, the contribution of
ultrashort distances at $I^+$ to the Planckian spectrum. The
analysis shows that, for  Schwarzschild astrophysical black holes,
the Hawking radiation (for both bosons and fermions) is very
robust up to very high frequencies (typically two orders above
Hawking's temperature). Below this scale, the contribution  of
ultrashort distances to the spectrum is negligible. We argue,
using a simple model with modified two-point functions, that the
above result seems to have a general validity and that it is
related to the observer independence of the short-distance
behavior of the corresponding two-point function. The above
suggests that only at high emission frequencies could an
underlying quantum theory of gravity potentially predict
significant deviations from Hawking's semiclassical result.
\end{abstract}


\maketitle

\section{Introduction}

Semiclassical gravity predicts the radiation of quanta by black
holes \cite{hawk1, parkerwald75}. The emission rate is given by
the product of the Planckian factor times the grey-body coefficient
$\Gamma_{lmp}(w)$\be \label{formula1}\frac{dN_{lmp}(w)}{dw dt}= \frac{1}{2\pi}
\Gamma_{l mp}(w)\frac{1}{e^{2\pi \kappa^{-1}(w-m\Omega_H
-q\Phi_H)} \pm 1} \ , \ee where $\kappa$, $\Omega_H$ and $\Phi_H$
are the surface gravity, angular velocity and the electric potential of the
black hole horizon. The signs $\pm$ in the denominator account for
the Bose or Fermi statistics, and $m$, $p$ and $q$ are the
corresponding axial angular momentum, helicity, and charge of the
radiated particle.\\
The deep connection of this result with thermodynamics
\cite{bardeen} and, in particular, with a generalized second law
\cite{bekenstein}  strongly supports its robustness \cite{waldbook,
frolov-novikov, icp2005}. However, as stressed in
\cite{jacobson9193}, a crucial ingredient in deriving Hawking
radiation via semiclassical gravity is the fact that any emitted
quanta, even those with very low frequency at future infinity,
will suffer a divergent blueshift when propagated backwards
in time and measured by a freely falling observer. Also,  in the
derivation of Fredenhagen and Haag \cite{fredenhagen-haag90}, the
role of the short-distance behavior of the two-point function is
fundamental. All derivations seem to invoke Planck-scale physics.
The exponential blueshift effect of the horizon of the
black hole could thus be regarded as a magnifying glass that makes
visible the ultrashort-distance physics to external observers.
According to this reasoning the microscopic structure
offered by string theory (or any other underlying theory) could
leave some imprint or signal in the emission rate. However, the
results of string theory seem to agree with  Hawking's prediction. For the
emission of low-energy quanta (with wavelength large compared to the
black hole radius), and for some particular near-extremal
charged black holes, the prediction of string theory
\cite{strings, reviews} coincides with the rate  (\ref{formula1}).
 This agreement is complete, despite the fact that the two calculations are very
different. For instance, whereas in semiclassical gravity one can
naturally split the emission rate into two factors (pure Planckian
black-body term and grey-body factor),  in the D-brane
derivation one gets directly the final answer without the above
mentioned splitting.\\

While the calculation of string theory requires low frequencies
for the emitted particles, in the arena of semiclassical gravity
the result is valid for all wavelengths, even those
smaller than the size of the black hole. The thermodynamic picture
strongly suggests the robustness of Hawking's prediction and its
interpretation as a low-energy effect, not affected by the
particular underlying theory of quantum gravity (see also
\cite{maldacena-strominger}), and expected to be valid for a large
range of frequencies. However, from the perspective of quantum
field theory in curved spacetime, it is unclear how to introduce a
cutoff in the scheme (parameterizing our ignorance on
trans-Planckian physics) in such a way that, for low-energy
emitted quanta, the decay rate (\ref{formula1}) is kept unaltered.
The aim of this work is to study this  issue in some
detail\footnote{We note that this sort of problem has already been
addressed in the context of acoustic black holes by modifying the
dispersion relation of the wave equation obeyed by sonic
disturbances \cite{unruh95, bmps-cj}. This is naturally justified
as an effect of the atomic microscopic structure of the fluid, and
requires a breakdown of Lorentz invariance. The rest frame of the
atoms of the fluid plays a privileged role. In this paper we
follow an alternative route.}.\\

In section \ref{sec:Bog} we will  review the standard derivation of black
hole radiation emphasizing the role of ultrahigh frequencies to
get the Planckian spectrum. In section \ref{sec:III} we rederive the Hawking
effect in terms of  two-point functions, instead of Bogolubov
transformations (for a general reference see \cite{parker-toms}),
  for both massless scalar and spin-$1/2$ fields.
 The new derivation of the black hole decay rate offers an
explicit way to evaluate the contribution of ultrashort
(Planck-scale) distances to the thermal Hawking spectrum. This is
the subject of section IV. In section V we present a simple model,
where the two-point functions are deformed with a Planck-length
parameter, to show how the previous results emerge in this new
scenario and support their robustness. We point out
that a generalized Hadamard condition plays a fundamental role to
keep unaltered the bulk of the Hawking effect. Finally, in section
VI, we summarize our conclusions and make some speculative
comments. In the appendices we give details of some calculations
used in the body of the text.

\section{Bogolubov coefficients and black hole radiance \label{sec:Bog}}

Let us consider the formation process of a Schwarzschild black
hole, as depicted in Fig.1, and a massless real scalar  field
$\phi$ propagating in this background. The equation of motion
obeyed by the field is $\Box \phi =0$ and the Klein-Gordon scalar
product is given by \be \label{KGscalarproduct} (\phi_1, \phi_2) =
-i\int_{\Sigma}d\Sigma^{\mu}(\phi_1\partial_{\mu}\phi_2^{*}-\phi_2^{*}\partial_{\mu}\phi_1)
\ , \ee where $\Sigma$ is a suitable ``initial data''
hypersurface. A natural choice for $\Sigma$ is the past null
infinity $I^-$ and therefore one can express the field in a set of
modes $u^{in}_j(x)$, which have positive frequency  in $I^-$ \be
\phi=\sum_i (a^{in}_i u^{in}_i+ a_i^{in \dagger} u_i^{in*})   \ .
\ee
\begin{figure}[htbp]
\begin{center}
\includegraphics[angle=0,width=2.2in,clip]{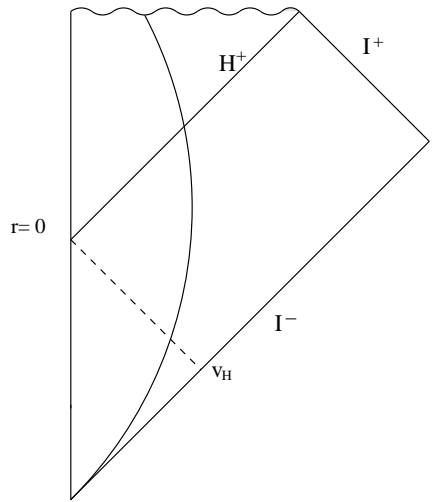}
\label{Fig1} \caption{Penrose diagram of a collapsing body
 producing a Schwarzschild black hole.}
\end{center}
\end{figure}
Alternatively, we can choose $\Sigma$ as $\Sigma = I^+ \cup H^+$,
where $I^+$ is the future null infinity and $H^+$ is the event
horizon. According to this we can then expand the field in an
orthonormal set of modes $u^{out}_i(x)$, which have positive
frequency with respect to the inertial time at $I^+$ and have zero
Cauchy data in $H^+$, together with a set of modes $u^{int}_i(x)$
with null outgoing component at $I^+$. Therefore we can write
 \be \phi =
\sum_{i}(a_i^{out} u^{out}_i+ a_i^{out \dagger} u_i^{out*})+
(a_i^{int} u^{int}_i+ a_i^{int \dagger} u_i^{int*})\ . \ee The
particular choice of modes $u^{int}_i$ does not affect the
computation of particle production at $I^+$, so we leave them
unspecified.

The modes $u^{out}_j(x)$ can be expressed in terms of the basis
$u^{in}_i$  \be u^{out}_j(x)=\sum_i \alpha _{ji}u^{in}_i(x)+\beta
_{ji}{u^{in*}_i(x)} \ , \ee where the coefficients $\alpha _{ji}$
and $\beta _{ji}$ are the so-called Bogolubov coefficients and are
given by the scalar products \be \alpha_{ij}=(u^{out}_i, u^{in}_j) \
, \ \ \beta_{ij}=-(u^{out}_i, u^{in*}_j) \ . \ee The above expansion
leads to an analogous relation for the creation and annihilation
operators:
 \be a_i^{out}=\sum_j (\alpha^*
_{ij}a_j^{in}-\beta ^*_{ij}a_j^{in \dagger}) \ . \ee When the
coefficients $\beta _{ij}$ do not vanish  the vacuum states
$|in\rangle$ and $|out\rangle$, defined as $ a_i^{in}|in
\rangle=0$ and $ a_i^{out}|out\rangle=0$, do not coincide and, as
a consequence, the number of particles measured in the $i^{th}$
mode by an ``out'' observer, $N_i^{out}=\hbar^{-1}a_i^{out
\dagger} a_i^{out}$, in the state $|in\rangle$ is given by
 \be \label{Niout}\langle in| N_i^{out}|in \rangle = \sum_k
|\beta_{ik}|^2 \ . \ee

Let us now briefly summarize the main steps of Hawking's
derivation. Assuming for simplicity that the background is
spherically symmetric we can  choose the  following  basis for the
ingoing and outgoing modes \be \label{uinlm}u^{in}_{wlm}|_{I^-}
\sim \frac{1}{ \sqrt{4\pi w}}\frac{e^{-iwv}}{r}Y_{l}^m(\theta,
\phi) \ , \ee \be \label{uoutlm}u^{out}_{wlm}|_{I^+} \sim
\frac{1}{ \sqrt{4\pi w}}\frac{e^{-iwu}}{r}Y_{l}^m(\theta, \phi) \
. \ee Here $Y_l^m(\theta, \phi)$ are the spherical harmonics. One
can evaluate the coefficients $\beta_{wlm,w'l'm'}$
 according to previous expressions by making the convenient choice
$\Sigma = I^-$ \be \beta_{wlm,w'l'm'}= i\int_{I^-} dvr^2d\Omega
(u^{out}_{wlm}\partial_{v}u^{in}_{w'l'm'}-
u^{in}_{w'l'm'}\partial_{v}u^{out}_{wlm}) \ . \ee The angular
integration is straightforward and leads to delta functions
$\delta_{l l'}$ $\delta_{m, -m'}$ for the $\beta$ coefficients. The
relevant point is to realize that the coefficients can be evaluated
and have a unique answer, which turns out to be independent of the
details of the collapse, if $u^{out}_i$ represents a late-time
wave-packet mode (i.e., centered around an instant $u$ with $u\to
+\infty$ along $I^+$). When these modes are propagated backwards in
time they are largely blueshifted when they approach the event
horizon. After  passing through the collapsing body they are
scattered  to $I^-$ in a very small interval just before $v_H$. To
know how they behave on $I^-$ (as needed to evaluate the scalar
product with $u^{in}_{wlm}$) one can apply the geometrical optics
approximation since the effective frequency, as measured by freely
falling observers, is very large. The (late-time) mode
$u^{out}_{wlm}$, which is of the form (\ref{uoutlm}) at $I^+$,
evolves and arrives at $I^-$ with the form \be
\label{uoutlmI-}u^{out}_{wlm}|_{I^-} \sim \frac{t_l(w)}{ \sqrt{4\pi
w}}\frac{e^{-iwu(v)}}{r}Y_{l}^m(\theta, \phi)\Theta(v_H-v) \ , \ee
where $t_l(w)$ is the transmission coefficient for the Schwarzschild
metric and the relation between null inertial coordinates $u$ at
$I^+$ and $v$ at $I^-$ is typically given by the logarithmic term
\be \label{relationuv0} u=v_H -\kappa^{-1}\ln{\kappa|v_H - v|} \ ,
\ee where, for the Schwarzschild black hole, $\kappa=1/4M$ and $v_H$
represents the location of the null ray that will form the event
horizon $H^+$ (see Fig. 1). One has then all ingredients to work out
the (late-time) Bogolubov coefficients \be \label{betaintegral}
\beta_{wlm,w'l'm'}=\frac{-(-)^m
t_l(w)}{2\pi}\sqrt{\frac{w'}{w}}\int_{-\infty}^{v_{H}}dve^{-iw(v_{H}-\kappa^{-1}\ln
\kappa|v_{H}-v|) -i w'v  } \delta_{ll'}\delta_{m\ -m'}\ . \ee They
can be  evaluated explicitly \be
\label{betaww'}\beta_{wlm,w'l'm'}=\frac{-(-)^mt_l(w)}{2\pi\kappa}\sqrt{\frac{w'}{w}}\
\frac{e^{-i(w+w')v_{H}}}{ (-\kappa^{-1}w'i+
\epsilon)^{1+\kappa^{-1}wi}}
\Gamma(1+\kappa^{-1}wi)\delta_{ll'}\delta_{m-m'} \ , \ee
 where we have introduced a negative real
part $(-\epsilon)$ into the exponent of (\ref{betaintegral}) to
ensure convergence of the corresponding integrals. To compute the
particle production at $I^+$ one has to evaluate
 the integral (from now on in this section we shall omit, for simplicity, the
subscripts $l, m$) \be \label{integralbeta}\int_{0}^{+\infty} dw'
\beta_{w_{1}w'} \beta_{w_{2}w'}^{*} \ . \ee The integration in
$w'$ reduces to \be \label{integralw'}\int_{0}^{+\infty}
\frac{dw'}{w'}e^{-\kappa^{-1}w_{1}i\ln(-\kappa^{-1}w'-i\epsilon)}
 e^{\kappa^{-1}w_{2}i \ln(\kappa^{-1}w'-i\epsilon)}=2\pi\kappa e^{-\pi \kappa^{-1}w_{1}} \delta (w_{1}-w_{2}) \ , \ee
from which we finally get \be \int_{0}^{+\infty} dw'
\beta_{w_{1}w'}\beta_{w_{2}w'}^{*}= \frac{|t_l(w_1)|^2}{e^{2\pi
\kappa^{-1}w_{1}} - 1} \delta (w_{1}-w_{2}) \ , \ee where the
coefficient in front of $\delta (w_{1}-w_{2})$
represents a steady thermal flow of radiation of frequency $w=w_1$
\be \label{planck} \frac{dN_{lm}(w)}{dw dt} \equiv
\frac{1}{2\pi}\langle in |N^{out}_{wlm}|in  \rangle=
\frac{1}{2\pi}\frac{\Gamma_l(w)}{e^{2\pi \kappa^{-1} w} - 1}\ ,
\ee and  the grey-body factor is given by $\Gamma_l(w)\equiv
|t_l(w)|^2$. For a generic collapse the result leads to
formula (\ref{formula1}).\\

It is important to remark at this point that a basic step to
exactly obtain the Planckian spectrum is (\ref{integralw'}), which
crucially requires an unbounded integration in all frequencies
$w'$. In fact, if we introduce an ultraviolet  cutoff $\Lambda$
for $w'$ we should replace (\ref{integralw'})  by \be \label{integralw'2}\int_{0}^{+\Lambda}
\frac{dw'}{w'}e^{-\kappa^{-1}w_{1}i\ln(-\kappa^{-1}w'-i\epsilon)}
 e^{\kappa^{-1}w_{2}i \ln(\kappa^{-1}w'-i\epsilon)}=e^{-\pi\kappa^{-1} w_{1}} 2\pi \delta_{\sigma}[\kappa^{-1}(w_1-w_2)] \ , \ee
where we have defined
\begin{eqnarray}
\delta_\sigma[\kappa^{-1}(w_1-w_2)]&=&\frac{\sin\left[\frac{\kappa^{-1}(w_1-w_2)}{\sigma}\right]}{\pi \kappa^{-1}(w_1-w_2)} \\
\sigma&=&\frac{1}{\ln [\kappa^{-1}\Lambda]}
\end{eqnarray}
Note that in the limit as $\sigma$ goes to zero $\delta_\sigma$
turns into Dirac's delta function and we recover
(\ref{integralw'}).
The new expression is, however, qualitatively different from the
previous one. To evaluate the new emission rate requires making
use of normalized wave-packet modes. Introducing the standard ones
\cite{hawk1}\be \label{eq:wavepacket}
u^{out}_{jnlm}=\frac{1}{\sqrt{\epsilon}}\int^{(j+1)\epsilon}_{j\epsilon}
dw\ e^{2\pi iw n/\epsilon}\ u^{out}_{w lm}\ ,  \ee where $j\geq 0$
and $n$ are integers,  representing wave-packets peaked around the
retarded time $u_n=2\pi n/\epsilon$
 and centered, with width $\epsilon$,
around the frequency $w_j\equiv (j+ 1/2)\epsilon $; and,
accordingly, defining\be
\beta_{jn,w'}=\frac{1}{\sqrt{\epsilon}}\int^{(j+1)\epsilon}_{j\epsilon}
dw\ e^{2\pi iw n/\epsilon}\ \beta_{ww'}\ , \ee the emission rate
results (see appendix A):
\begin{equation}\label{eq:N-timedependent}
\langle in |N^{out,\sigma}_{jn}|in \rangle \approx
\frac{|t_l(w_j)|^2}{e^{2\pi\kappa^{-1}w_j}-1}\frac{\sin\left[\left(\frac{2\pi
n}{\epsilon}-v_H\right)\frac{\pi\kappa\sigma}{2}\right]}{\left[\left(\frac{2\pi
n}{\epsilon}-v_H\right)\frac{\pi\kappa\sigma}{2}\right]}
\end{equation}
From this expression\footnote{We note that the oscillatory
behavior in (\ref{eq:N-timedependent}) is an artifact of the particular way
we have introduced the cutoff. If the  cutoff is introduced in a
different way, see appendix A, the oscillatory term disappears but the  decay with time is maintained
as $\sim e^{-[(2\pi n/\epsilon-v_H)(\pi\kappa\sigma/2)]^2}$.} we see that the rate of emitted particles
depends on the retarded time $u_n= 2\pi n/\epsilon$ and decays
with time for any small but nonzero value of
$\sigma=1/\ln[\Lambda/\kappa]$. Only when
$\Lambda$ goes to infinity (no high frequency cutoff) do we recover
the steady thermal flux of radiation.\\

In conclusion, the above discussion shows that the radiation is
now time-dependent  and decays for all finite values of $\Lambda$.
The decay in time would also occur at low frequencies, where string
theory agrees with Hawking's prediction. Therefore, as expected from
conventional arguments, the mathematical role of the ultrahigh
 frequencies is very important for the late-time behavior.
 Nevertheless, since they only enter as virtual quanta, their
actual status is unclear \cite{Parker77}. A derivation of the
Hawking effect  based on quantities defined on the asymptotically
flat region, where physical observations are made, would be preferable. This turns out
to be possible if, instead of working with Bogolubov coefficients,
one uses two-point functions. They are defined in the $I^+$ region
where a Planck-length cutoff in ``distances'' can be
naturally introduced. This is the task of next sections.

\section{Two-point functions and  black hole radiance \label{sec:III}}

This section will be devoted to rederive Hawking radiation by
means of two-point functions. Intuitively the idea is simple. In
the conventional analysis in terms of Bogolubov coefficients, we
first perform the integration in distances (to compute the scalar
product required for the $\beta$ coefficients) and leave to the
end the integration in frequencies $w'$. In contrast, we can
invert the order and perform first the integration in frequencies
(which naturally leads to introduce the two-point function of the
matter field) and perform the integration in distances at the end.

Let us rewrite the basic expression (\ref{Niout}), or more
precisely, the expectation values of the operator
$N_{i_1i_2}^{out}\equiv \hbar^{-1}\ {a^{out}}^\dagger_{i_1}
a^{out}_{i_2} $, as follows \bea \label{Nijout} &&\langle
in|N_{i_1i_2}^{out}|in\rangle =\sum_k
\beta_{i_1k}\beta_{i_2k}^*=-\sum_k
(u^{out}_{i_1}, u^{in*}_k)(u^{out*}_{i_2}, u^{in}_k)=\nonumber \\
&=&\sum_k\left(\int_{\Sigma}d\Sigma^{\mu}_1u^{out}_{i_1}(x_1)
{\buildrel\leftrightarrow\over{\partial}}_\mu u^{in}_k(x_1)\right)
\left(\int_{\Sigma}d\Sigma^{\nu}_2u^{out
*}_{i_2}(x_2){\buildrel\leftrightarrow\over{\partial}}_\nu u^{in
*}_k(x_2)\right)\ . \ \eea If we now consider the sum in modes
before making the integrals of the two scalar products, and take
into account that \be \langle in| \phi (x_1)\phi (x_2)| in
\rangle=\hbar\sum_k u_k^{in}(x_1){u_k^{in}}^*(x_2) \ , \ee we
obtain a simple expression for the particle production number in
terms of the two-point function
\begin{equation}\label{eq:N-eps}
\langle in|N_{i_1i_2}^{out}|in\rangle = \hbar^{-1} \int_\Sigma
d\Sigma_1 ^\mu d\Sigma_2 ^\nu
[u^{out}_{i_1}(x_1){\buildrel\leftrightarrow\over{\partial}}_\mu
][u^{out*}_{i_2}(x_2){\buildrel\leftrightarrow\over{\partial}}_\nu
]\langle in| \phi (x_1)\phi (x_2)|in\rangle \ .
\end{equation}

In the above expression the two-point function should be then
interpreted in the distributional sense. The
``$i\epsilon$-prescription'' (see eq.(\ref{N-epsI-})
below) is therefore assumed for the two-point distribution
$\langle in| \phi (x_1)\phi (x_2)|in\rangle$ and it verifies the
Hadamard condition\footnote{The two-point distribution should have,
for all physical states, a short-distance structure similar to
that of the ordinary vacuum state in Minkowski space:
$(2\pi)^{-2}(\sigma +2i\epsilon t + \epsilon^2)^{-1}$, where
$\sigma(x_1,x_2)$ is the squared geodesic distance.}
\cite{waldbook,Kay-Wald}. Alternatively, taking into account the
trivial identity $\langle out|{a^{out}}^\dagger_{i_1}
a^{out}_{i_2}|out\rangle =0$ we can rewrite the above expression
as \cite{agullo-navarro-salas-olmo06}
 \be \label{eq:N-nord}
\langle in|N_{i_1i_2}^{out}|in\rangle = \hbar^{-1} \int_\Sigma
d\Sigma_1 ^\mu d\Sigma_2 ^\nu
[u^{out}_{i_1}(x_1){\buildrel\leftrightarrow\over{\partial}}_\mu
][u^{out*}_{i_2}(x_2){\buildrel\leftrightarrow\over{\partial}}_\nu
]\langle in| :\phi (x_1)\phi (x_2):|in\rangle \ , \end{equation}
where $\langle in| :\phi (x_1)\phi (x_2):|in\rangle \equiv \langle
in| \phi (x_1)\phi (x_2)|in\rangle - \langle out| \phi (x_1)\phi
(x_2)|out\rangle$. Now the Hadamard condition for both $|in\rangle$
and $|out\rangle$ states ensures that $\langle in| :\phi (x_1)\phi
(x_2):|in\rangle$ is a smooth function.

\subsection{Thermal spectrum for a scalar field}

Let us now apply this scheme in  the formation process of a
spherically symmetric  black hole and restrict the ``out'' region to
$I^+$. The ``in'' region is, as usual, defined by $I^-$. At $I^+$ we
can consider the conventional radial plane-wave modes \be
u^{out}_{wlm}(t,r, \theta, \phi)|_{I^+}\sim
u^{out}_{w}(u)\frac{Y_{l}^m(\theta, \phi)}{r} \ , \ee where
$u^{out}_{w}(u)= \frac{e^{-iwu}}{\sqrt{4\pi w}} $. We shall now
evaluate the matrix coefficients $\langle
in|N_{i_1i_2}^{out}|in\rangle$ where
$i_{1,2}\equiv(w_{1,2},l_{1,2},m_{1,2})$. Taking as the initial
value hypersurface $I^-$ and integrating by parts
 we obtain
\bea \label{Nblackhole} &&\langle in|N_{i_1i_2}^{out}|in \rangle =
\frac{4}{\hbar}\int_{I^-}r_1^2dv_1d\Omega_1\int_{I^-}r_2^2dv_2d\Omega_2
u^{out}_{w_1} u^{out*}_{w_2} \times \nonumber
\\ &&\frac{Y_{l_1}^{m_1}(\theta_1,\phi_1)}{r_1}\frac{Y_{l_2}^{m_2*}(\theta_2,\phi_2)}{r_2}\partial_{v_1}\partial_{v_2}
\langle in|\phi(x_1)\phi(x_2)|in\rangle \ . \eea The two-point
function above can be now expanded at $I^-$ as \be
\label{correlatorin}\langle in|\phi(x_1)\phi(x_2)|in\rangle=\hbar
\int_0^{\infty}dw\sum_{l,m}\frac{e^{-iwv_1}}{\sqrt{4\pi
w}}\frac{Y_{l}^m(\theta_1,\phi_1)}{r_1}\frac{e^{iwv_2}}{\sqrt{4\pi
w}}\frac{Y_{l}^{m*}(\theta_2,\phi_2)}{r_2} \ . \ee  Recall that
the radial part of the late-time ``out'' modes $u^{out}_{wlm}$,
when they are propagated backward in time and reach $I^-$, takes
the form \be \label{propagation} u^{out}_{w}|_{I^-}\sim
t_l(w)\frac{e^{-iwu(v)}}{\sqrt{4\pi w}} \Theta(v_H-v)\ee where
$u(v)\approx v_H-\kappa^{-1}\ln \kappa(v_H-v)$.
Performing now angular integrations and taking into account that
\be \label{shortdistance}
\partial_{v_1}\partial_{v_2}\int_0^{\infty}dw\frac{e^{-iw(v_1-v_2)}}{4\pi
w}=-\frac{1}{4\pi}\frac{1}{(v_1-v_2-i\epsilon)^2} \ , \ee
we get
\bea \label{N-epsI-}\langle in|N_{i_1i_2}^{out}|in \rangle
= -\frac{t_{l_1}(w_1)t^*_{l_2}(w_2)}{4\pi^2\sqrt{w_1w_2}}\int_{-\infty}^{v_{H}}dv_1 dv_2
\frac{e^{-iw_1u(v_1)+iw_2u(v_2)}}{(v_1-v_2-i\epsilon)^2} \delta_{l_1
l_2}\delta_{m_1 m_2}\ , \eea  where the limit $\epsilon \to 0^+$ is
understood. Alternatively, since we are interested in quantities measured at $\Sigma=I^+$, we could use this latter hypersurface to carry out the calculations. In this case, the expression for the particle production rate becomes
\bea \label{N-epsI+}\langle in|N_{i_1i_2}^{out}|in \rangle
= -\frac{t_{l_1}(w_1)t^*_{l_2}(w_2)}{4\pi^2\sqrt{w_1w_2}}\int_{-\infty}^{\infty}du_1 du_2
\frac{\frac{dv}{du}(u_1)\frac{dv}{du}(u_2)}{[v(u_1)-v(u_2)-i\epsilon]^2}e^{-iw_1u_1+iw_2u_1} \delta_{l_1
l_2}\delta_{m_1 m_2}\ , \eea
and leads to
 \bea
\label{Nregulatedepsilon} \langle in|N_{i_1i_2}^{out}|in \rangle =
 \frac{-t_{l_1}(w_1)t^*_{l_2}(w_2)}{4\pi^2\sqrt{w_1w_2}}\int_{-\infty}^{+\infty}du_1 du_2
\frac{(\frac{\kappa}{2})^2e^{-iw_1u_1+iw_2u_2}}{[\sinh
\frac{\kappa}{2}(u_1-u_2-i\epsilon)]^2} \delta_{l_1
l_2}\delta_{m_1 m_2}\ . \eea This last expression is more convenient for computational purposes. Since the function in the integral depends only on the difference $z\equiv u_2-u_1$, the integral in $u_2+u_1$ can be performed immediately and leads to a delta function in frequencies
\begin{equation}\label{eq:divergencesboson}
\langle in|N_{i_1 i_2}^{out}|in\rangle =
-\frac{t_{l_1}(w_1)t^*_{l_2}(w_2)\delta
(w_1-w_2)}{2\pi\sqrt{w_1w_2}} \int_{-\infty}^{+\infty} dz
e^{-i\frac{w_1+w_2}{2}z} \frac{(\frac{\kappa}{2})^2\delta_{l_1
l_2}\delta_{m_1 m_2}}{[\sinh \frac{\kappa}{2}(z -i\epsilon)]^2} \ ,
\end{equation}
Performing the integration in $z$
we recover the Planckian spectrum and the particle production rate
\be \langle in |N^{out}_{wlm}|in \rangle=
\frac{|t_l(w)|^2}{e^{2\pi \kappa^{-1} w} - 1}\ . \ee

This derivation of black hole radiation is somewhat
parallel to the one given in \cite{fredenhagen-haag90}. The
emphasis is in the  two-point function of the quantum
state, instead of the usual treatment in terms of Bogolubov
transformations. It is worth noting that (\ref{N-epsI-})
displays an apparent sensitivity to ultrashort distances
 due to the highly oscillatory
behavior of the modes in a small region before $v_H$. A similar
conclusion can be obtained from (\ref{eq:divergencesboson}) when
$z\to 0$. The sensitivity to short distances is, however, less
apparent if we repeat the above calculations using the expression
(\ref{eq:N-nord}) instead of (\ref{eq:N-eps}). In this case, we
find \bea \label{N-nordI-}\langle in|N_{i_1i_2}^{out}|in \rangle
&=&
 -\frac{t_{l_1}(w_1)t^*_{l_2}(w_2)}{4\pi^2\sqrt{w_1w_2}}\int_{-\infty}^{v_H}
 dv_1 dv_2
e^{-iw_1u(v_1)+iw_2u(v_2)}\times \nonumber
\\&&\left [\frac{1}{(v_1-v_2)^2}-\frac{\frac{du}{dv}(v_1)\frac{du}{dv}(v_2)}{[u(v_1)-u(v_2)]^2} \right ] \delta_{l_1
l_2}\delta_{m_1 m_2}\ , \eea where we have dropped the $i\epsilon$
terms since they are now redundant. Note that  the short-distance
divergence of  $1/(v_1 -v_2)^2$ in (\ref{N-nordI-}) is exactly
cancelled by
$\frac{\frac{du}{dv}(v_1)\frac{du}{dv}(v_2)}{[u(v_1)-u(v_2)]^2}$
for any smooth choice of the function $u(v)$. This cancellation is
a consequence of the Hadamard condition that verify both ``in''
and ``out'' vacuum states. It is also important to remark that the
above formula exhibits the absence of particle production under
conformal-type (Möbius) transformations \be v=\frac{au +b}{cu +d}
\, \ee
where $ab-cd=1$.\footnote{ This leads, immediately, to the
expected result that there is no particle production under Lorentz
transformations.}

If the calculation is performed using $\Sigma = I^+$, one finds
\bea \label{N-nordI+} \langle in|N_{i_1i_2}^{out}|in \rangle &=&
 -\frac{t_{l_1}(w_1)t^*_{l_2}(w_2)}{4\pi^2\sqrt{w_1w_2}}\int_{I^+}
 du_1 du_2
e^{-iw_1u_1+iw_2u_2}\times \nonumber
\\&&\left [\frac{\frac{dv}{du}(u_1)\frac{dv}{du}(u_2)}{(v(u_1)-v(u_2))^2}-\frac{1}{[u_1-u_2]^2} \right ] \delta_{l_1
l_2}\delta_{m_1 m_2}\ , \eea which leads to
\bea\label{eq:divergences2} \langle in|N_{i_1 i_2}^{out}|in\rangle
&=& -\frac{t_{l_1}(w_1)t^*_{l_2}(w_2)\delta
(w_1-w_2)}{2\pi\sqrt{w_1w_2}} \int_{-\infty}^{+\infty} dz
e^{-i\frac{w_1+w_2}{2}z}\times \nonumber \\ && \left
[\frac{(\frac{\kappa}{2})^2}{(\sinh \frac{\kappa}{2}
z)^2}-\frac{1}{z^2}\right ] \delta_{l_1 l_2}\delta_{m_1 m_2} \ .
\eea  The integral in  distances $z$ also leads, as expected, to the Hawking
formula\footnote{For Kerr-Newman black holes the calculation is
similar up to a shift in the wave function $e^{-iwz}$, which
should be now replaced by $e^{-i(w-m\Omega_{H}-q\Phi_H)z} $, as an
effect of wave propagation through the corresponding potential
barrier.} \bea \label{Norderingfinal0}\langle in |N^{out}_{wlm}|in
\rangle &=& -\frac{|t_l(w)|^2}{2\pi w} \int_{-\infty}^{+\infty} dz
e^{-iwz} \left[\frac{(\frac{\kappa}{2})^2}{(\sinh \frac{\kappa}{2}
z)^2}-\frac{1}{z^2}\right ]\nonumber
\\ &=& \frac{|t_l(w)|^2}{e^{2\pi w \kappa^{-1}}-1} \ . \eea

\subsection{Thermal spectrum for a $s=1/2$ field}
In this subsection we shall extend the analysis of the scalar
field to a fermionic $s=1/2$ field. For simplicity we take a
massless Dirac field, obeying the wave equation \be
\gamma^{\mu}\nabla_{\mu} \psi =0 \ , \ee where $\gamma^{\mu} =
V^{\mu}_{a} \gamma^{a}$ are the curved space counterparts of the
Dirac matrices $\gamma^{a}$ (see appendix B for calculations
omitted in this section). The Klein-Gordon scalar product
(\ref{KGscalarproduct}) is now replaced by \be (\psi_1, \psi_{2})=
\int_{\Sigma} d\Sigma^{\mu}\bar{\psi}_1 \gamma_{\mu}\psi_2 \ . \ee
Therefore the expression for the expectation values
(\ref{eq:N-eps}) is replaced by
\begin{equation}\label{eq:ai+ajspinor}
\langle in|N^{out}_{i_1i_2}|in\rangle = \hbar^{-1} \int_\Sigma
d\Sigma_1 ^\mu d\Sigma_2 ^\nu  [\bar{u}^{out }_{i_2}(x_2)
{\gamma}_\nu]_{b}[\gamma_{\mu}u^{out}_{i_1}(x_1)]^{a}\langle in|
\bar{\psi}_a (x_1)\psi^b(x_2)|in\rangle \ .
\end{equation}
At $I^+$ we can consider the  normalized radial plane-wave
modes\footnote{On physical grounds we should use left-handed
spinors $u^{out}_{L,w j m_j}\equiv\frac{1}{\sqrt{2}}(u^{out}_{w \
|\kappa_j| m_j} - u^{out}_{w \ -|\kappa_j| m_j})$.
The final result is not changed. See appendix B.} \bea
u^{out}_{w\kappa_j m_j}(t,r, \theta, \phi)|_{I^+}\sim
\frac{e^{-iwu}}{\sqrt{4\pi}r}\left(\begin{array}{c}
                               \eta(\hat{r})^{m_j}_{\kappa_j}\\
                                (\hat{r}\vec{\sigma})\eta(\hat{r})^{m_j}_{\kappa_j}
                                \end{array}\right)
 \ , \eea where $\eta(\hat{r})^{m_j}_{\kappa_j}$ are
two-component spinor harmonics (see appendix B). Note that the
angular momentum quantum number $j$ is uniquely determined by the
relation $\kappa_j=\pm(j+1/2)$. The above modes, when propagated
backwards in time and reach $I^-$, turn into
\begin{eqnarray}\label{eq:OUT-mode}
u^{out}_{w\kappa_j m_j}(t,r, \theta, \phi)|_{I^-}\sim
t_{\kappa_j}(w)\frac{e^{-iwu(v)}}{\sqrt{4\pi}r}\left(\begin{array}{c}
                               \eta(\hat{r})^{m_j}_{\kappa_j}\\
                                -(\hat{r}\vec{\sigma})\eta(\hat{r})^{m_j}_{\kappa_j}
                                \end{array}\right)\Theta(v_H-v)\sqrt{\frac{du(v)}{dv}}
\ , \end{eqnarray} where the last term $\sqrt{du(v)/dv}$ appears
due to the fermionic character of the field\footnote{The vierbein
fields needed to properly write the field equation in a curved
spacetime have been trivially fixed in the asymptotic flat
regions, so its transformation law under change of coordinates is
then translated to the spinor itself (see also appendix B).}.
Proceeding as in the bosonic case we can expand the two-point
function as
\begin{equation}\label{eq:2Pspinor}
\langle in| \bar{\psi}_a (x_1)\psi^b(x_2)|in\rangle= \hbar
\sum_k \bar{v}^{in}_{k,a}(x_1) v^{in,b}_k(x_2)
\end{equation}
where $v^{in}_k$ are negative-energy solutions which in $I^-$ take
the form \bea \label{eq:IN-mode} v^{in}_k\to
v^{in}_{w\kappa_jm}(t,r, \theta, \phi)|_{I^-}\sim
\frac{e^{iwv}}{\sqrt{4\pi}r}\left(\begin{array}{c}
                               \eta(\hat{r})^{m_j}_{\kappa_j}\\
                                -(\hat{r}\vec{\sigma})\eta(\hat{r})^{m_j}_{\kappa_j}
                                \end{array}\right)
 \ , \eea
Performing first the angular integrations and taking into account
the orthonormality relations of the spinor harmonics
$\eta(\hat{r})^{m_j}_{\kappa_j}$, the above formulas get
simplified and become
\begin{eqnarray}\label{eq:N-fermions}
\langle in|N_{i_1 i_2}^{out}|in\rangle &=&
-i\frac{t_{\kappa_{j_1}}(w_{1})t_{\kappa_{j_2}}^*(w_{2})}{4\pi^2}\delta_{m_{j_1}m_{j_2}}\delta_{\kappa_{j_1}\kappa_{j_2}}\times\nonumber\\
&\times&\int_{-\infty}^{v_H} dv_1
dv_2\sqrt{\frac{du(v_1)}{dv}\frac{du(v_2)}{dv}}\frac{e^{-iw_{1}u(v_1)+iw_{2}u(v_2)}}{(v_1-v_2-i\epsilon)}
\end{eqnarray}
As in the bosonic case, we rewrite this expression as an integral
over $I^+$
\begin{eqnarray}
\langle in|N_{i_1 i_2}^{out}|in\rangle &=&
-i\frac{t_{\kappa_{j_1}}(w_{1})t_{\kappa_{j_2}}^*(w_{2})}{4\pi^2}\delta_{m_{j_1}m_{j_2}}
\delta_{\kappa_{j_1}\kappa_{j_2}}\times\nonumber\\
&\times&\int_{-\infty}^{\infty} du_1
du_2e^{-iw_{1}u_1+iw_{2}u_2}\frac{(\frac{\kappa}{2})
}{\sinh[\frac{\kappa}{2}(u_1-u_2-i\epsilon)]}
\end{eqnarray}
We can also split the integral in a product of a function
dependent on $u_2+u_1$ and another function which depends on
$z\equiv u_2-u_1$. The former leads to a delta function in frequencies
and we are left with
\begin{eqnarray}\label{eq:divergences}
\langle in|N_{i_1 i_2}^{out}|in\rangle &=&
-\frac{i}{2\pi}t_{\kappa_{j_1}}(w_{1})t_{\kappa_{j_2}}^*(w_{2})\delta_{m_{j_1}m_{j_2}}\delta_{\kappa_{j_1}\kappa_{j_2}}\delta{(w_1-w_2)}\times\nonumber\\
&\times& \int_{-\infty}^{+\infty} dz e^{-i\frac{w_1+w_2}{2}z}
\frac{(\frac{\kappa}{2})}{\sinh [\frac{\kappa}{2}(z -i\epsilon)]}
\ ,
\end{eqnarray}
Performing now the integration in $z$  \be
\label{eq:divergences3}\frac{-i}{2\pi} \int_{-\infty}^{+\infty} dz
e^{-iwz} \frac{(\frac{\kappa}{2}) }{\sinh [\frac{\kappa}{2}(z
-i\epsilon)]} =\frac{1}{e^{2\pi w\kappa^{-1}}+1} \ , \ee  we
recover the Planckian spectrum, with the Dirac-Fermi statistics,
and the corresponding particle production rate \be \langle in
|N^{out}_{wm_j\kappa_j}|in \rangle=
\frac{|t_{\kappa_j}(w)|^2}{e^{2\pi \kappa^{-1} w} + 1}\ . \ee

Analogously as in the bosonic case, if we use the normal-ordering
prescription instead of the $i\epsilon$ one we find  \bea \langle
in|N_{i_1i_2}^{out}|in \rangle &=&
-i\frac{t_{\kappa_{j_1}}(w_1)t_{\kappa_{j_2}}(w_2)^*}{4\pi^2}\int_{-\infty}^{v_H}
 dv_1 dv_2 \sqrt{\frac{du(v_1)}{dv}\frac{du(v_2)}{dv}}\times
\\&&e^{-iw_1u(v_1)+iw_2u(v_2)}\left [\frac{1}{[v_1-v_2]}-\frac{\sqrt{\frac{du(v_1)}{dv}\frac{du(v_2)}{dv}}}{[u(v_1)-u(v_2)]} \right ] \delta_{\kappa_{j_1}
\kappa_{j_2}}\delta_{m_{j_1} m_{j_2}}\ \nonumber. \eea
Changing the integration surface to $I^+$ we get
\bea \langle in|N_{i_1i_2}^{out}|in
\rangle &=&
 -i\frac{t_{\kappa_{j_1}}(w_1)t_{\kappa_{j_2}}(w_2)^*}{4\pi^2}\int_{I^+}
 du_1 du_2 \sqrt{\frac{dv(u_1)}{du}\frac{dv(u_2)}{du}}\times
\\&&e^{-iw_1u_1+iw_2u_2}\left [\frac{\sqrt{\frac{dv(u_1)}{du}\frac{dv(u_2)}{du}}}{[v(u_1)-v(u_2)]}-\frac{1}{[u_1-u_2]} \right ] \delta_{\kappa_{j_1}
\kappa_{j_2}}\delta_{m_{j_1} m_{j_2}}\ \nonumber. \eea
Note again that the short-distance divergence of the two-point function of the ``out'' state $\frac{1}{[u_1-u_2]}$
is exactly cancelled by the corresponding one of the ``in'' state
$\frac{\sqrt{\frac{dv(u_1)}{du}\frac{dv(u_2)}{du}}}{[v(u_1)-v(u_2)]}$,
since both vacua are Hadamard states. After some algebra we get
\bea \label{Norderingfinal}\langle in |N^{out}_{w\kappa_jm_j}|in
\rangle= -i\frac{|t_{\kappa_j}(w)|^2}{2\pi}
\int_{-\infty}^{+\infty} dz e^{-iwz}
\left[\frac{(\frac{\kappa}{2}) }{\sinh (\frac{\kappa}{2}
z)}-\frac{1}{z}\right] \ , \eea and taking into account \be
\frac{-i}{2\pi } \int_{-\infty}^{+\infty} dz e^{-iwz}
\left[\frac{(\frac{\kappa}{2})}{\sinh (\frac{\kappa}{2}
z)}-\frac{1}{z}\right]= \frac{1}{e^{2\pi w \kappa^{-1}}+1} \ , \ee
we newly recover the fermionic thermal spectrum.

\section{Short-distance contribution to the Planckian spectrum \label{sec:short}}

\subsection{Bosons}
We have seen in the previous section that it is possible to
rederive the Hawking effect in terms of two-point functions.
Either via expressions (\ref{eq:N-eps}),
(\ref{N-epsI+}) or, equivalently, via expressions
(\ref{eq:N-nord}),(\ref{N-nordI+}). Both
prescriptions are equivalent and lead to the Planckian spectrum
modulated by  grey-body factors. The advantage of the final
expression (\ref{Norderingfinal0}) is that it offers an explicit
evaluation of the contribution of distances to the Planckian
spectrum. To be more explicit, the integral\footnote{\label{foot-x} We have intentionally omitted the grey-body factors $|t_l(w)|^2$ in (\ref{proposalI}) because they are irrelevant for the discussion of this section.}
 \be \label{proposalI}I^B(w\kappa^{-1}, \alpha
\kappa)=-\frac{1}{2\pi w} \int_{-\alpha}^{+\alpha} dz e^{-iwz}
\left[\frac{(\frac{\kappa}{2})^2}{(\sinh \frac{\kappa}{2}
z)^2}-\frac{1}{z^2}\right ] \ , \ee can be interpreted as the
contribution of  short-distances $z \in [-\alpha, \alpha]$ to the (bosonic)
thermal spectrum when $\alpha$ is close to the Planck length $l_P$.
One could, alternatively, be tempted to propose,
according to (\ref{N-epsI+}), the integral \be
I^B_{i\epsilon}(w\kappa^{-1}, \alpha \kappa)\equiv\frac{-1}{2\pi w}
\int_{-\alpha}^{+\alpha} dz e^{-iwz}
\frac{(\frac{\kappa}{2})^2}{[\sinh \frac{\kappa}{2}
(z-i\epsilon)]^2} \, \ee
 as a legitimate expression to account for
the short-distance contributions. However, this interpretation is
not physically sound. In the absence of a black hole, when there
is no radiation at all, the above expression becomes \be
\frac{-1}{2\pi w} \int_{-\alpha}^{+\alpha} dz e^{-iwz} \frac{1}{
(z - i\epsilon)^2} \ . \ee For $\alpha \to +\infty$ this
expression vanishes, as expected due to the absence of radiation.
However, for finite $\alpha$ it is non-vanishing.
In contrast, the proposed expression (\ref{proposalI}), does not
suffer from this weird behavior,  due to the presence of the
second term. \\

 In conclusion, the calculation of black hole radiation using the
prescription (\ref{eq:N-nord}) offers the possibility to
 re-evaluate Hawking radiation by removing the range of
distances where physics can be dominated by an underlying theory
beyond field theory.
We shall now work out explicitly the short-distance contribution to
Hawking radiation to see whether it is fundamental or not in order
to obtain the thermal spectrum. The integral (\ref{proposalI}) can
be worked out analytically
\begin{eqnarray} \label{primitivafranja}
 & & I^B(w\kappa^{-1},\alpha\kappa)= -\frac{Si(\alpha w)}{\pi}
-\frac{\kappa}{4 \pi w} \{e^{i \alpha w} (F[1,-i
w\kappa^{-1},1-iw\kappa^{-1},e^{-\alpha\kappa}] \ \ \ \ \ \ \ \ \
\ \ \ \ \nonumber
\\& & -F[1,i w\kappa^{-1},1+iw\kappa^{-1},e^{\alpha\kappa}])+e^{-i
\alpha w} (F[1,i
w\kappa^{-1},1+iw\kappa^{-1},e^{-\alpha\kappa}]\nonumber
\\ & &-F[1,-i w\kappa^{-1},1-iw\kappa^{-1},e^{\alpha\kappa}])\}+ \frac{1}{2
\pi \alpha w} \cos(\alpha w) \left[\alpha\kappa
\frac{(1+e^{\alpha\kappa})}{(e^{\alpha\kappa}-1)}-2\right]
\end{eqnarray}
where $F$ is a hypergeometric function and
$Si(x)=\int^x_{0}dt\frac{\sin t}{t}$. To get some insight about the
properties of this formula, we find useful to expand it in powers of
$w\kappa^{-1}$ and $\alpha\kappa$. The expansion in $\alpha\kappa$
assumes that the microscopic length scale $\alpha \sim l_P$ is much
smaller than the typical emission wavelength $\sim \kappa^{-1}$ of
the black hole, whose (macroscopic) temperature is
$T_H=\kappa/2\pi$. For a Solar-mass black hole $\alpha\kappa \sim
10^{-40}$ and for a  primordial black hole of $10^{15}$ g
$\alpha\kappa \sim 10^{-21}$. The expansion in $w\kappa^{-1}$ means
that we are looking at frequencies below the typical emission
frequency, $w_{typical}\sim
 T_H$,  of the black hole. The result is as follows
\bea
\label{expansionIB} I^B(w\kappa^{-1}, \alpha\kappa)&=& \left(
\frac{1}{12 \pi} \alpha\kappa-\frac{1}{720
\pi}(\alpha\kappa)^3+O[(\alpha\kappa)^5]\right)
\frac{\kappa}{w}\nonumber
\\ &-&\left(\frac{1 }{72 \pi}(\alpha\kappa)^3+O[(\alpha\kappa)^5]\right)
\frac{w}{\kappa} +\left(O[(\alpha\kappa)^5]\right)
(\frac{w}{\kappa})^3+\ldots \ \ \eea From this expansion we conclude
that  the contribution of short distances to the spectrum is
completely negligible in the very low energy regime $w/\kappa\ll 1$
since\be \label{eq:ratio1} \lim_{w\kappa^{-1} \to
0}\frac{I^B(w\kappa^{-1},\alpha\kappa)}{(e^{2\pi
w\kappa^{-1}}-1)^{-1}}=\frac{\alpha\kappa}{6} \ll1 \ . \ee Moreover,
due to the smallness of $\alpha\kappa$, we find that
$I^B(w_{typical}\kappa^{-1},\kappa\alpha)$ can be well approximated
by (\ref{expansionIB}) even for frequencies close to the typical
emission frequency, which leads to
\be\label{eq:ratio2}\frac{I^B(w_{typical}\kappa^{-1},\alpha\kappa)}{(e^{2\pi
w_{typical}\kappa^{-1}}-1)^{-1}} \sim 0.3 \ \alpha\kappa \ll 1 \ .
\ee Again, since $\alpha\kappa \ll 1$, we find a negligible
contribution at $w_{typical}\sim T_H$. To be precise, for a
Schwarzschild black hole of three solar masses, when $\alpha$ is
around the Planck length $l_P=1.6\times 10^{-33}cm $, the relative
contribution to the Planckian distribution $\frac{I^B(w\kappa^{-1},
\alpha\kappa)}{(e^{2\pi w\kappa^{-1}}-1)^{-1}}$ is, for
$w=w_{typical}$, of order $10^{-38}\%$. For primordial black holes,
$M\sim 10^{15}$ g, the relative contribution is still insignificant:
$10^{-19}\%$. Even more, using the expansion (\ref{expansionIB}) we
easily get
 \be \label{eq:ratio3}\frac{I^B(w\kappa^{-1},\alpha\kappa)}{(e^{2\pi
w\kappa^{-1}}-1)^{-1}} \approx \frac{\alpha\kappa (e^{2\pi
w\kappa^{-1}}-1)}{12\pi w\kappa^{-1}}\ , \ee and we find that, for
a black hole of three solar masses, we need to look at the high
frequency region, $w/w_{typical}\approx 96$, to find that the
contribution of Planck distances $I^B(w\kappa^{-1}, l_P\kappa)$ is
of order of the total spectrum itself\footnote{The exponential
behavior in frequencies  of the ratio (\ref{eq:ratio3}) explains
why potential deviations from thermality arise at frequencies much
lower than $w\sim 1/l_P$.} . For primordial black holes we find
$w/w_{typical}\approx 52$. The same numerical estimates can be
found using the exact analytical expressions.\\

We can also naturally ask about the contribution to the spectrum
of large distances. This question is immediately answered using
our analytical expression (\ref{primitivafranja}). The
contribution of distances up to $\alpha = 20r_g$, where $r_g$ is
the gravitational radius, represents 90\% of the thermal peak at
$w_{typical}$. For $\alpha =200 r_g$ we obtain 99.7\% and for
$\alpha=2\times 10^4 r_g$ the percentage is around 99.99998\% .
\\

Summarizing, we have provided a quantitative estimate of how much
of Hawking radiation is actually due to Planckian distances. It
turns out that the contribution of ultrashort distances is
negligible and thermal radiation is very robust up to frequencies
of order $96T_H$ (for Schwarzschild black holes of three solar
masses) or $52T_H$ (for primordial black holes). In parallel and
dual to this, the contribution of  large distances is also
insignificant.\\

It is interesting to repeat the same calculations with the
$i\epsilon$-prescription. As we have already stressed with this
prescription one cannot expect a meaningful result. The
 outcome is completely different. The contribution of distances
in the interval $z \in [-\alpha, +\alpha]$ is now \bea
I^B_{i\epsilon\rightarrow0}(w\kappa^{-1},\alpha\kappa)&=&
\frac{e^{\alpha\kappa(1-iw\kappa^{-1})}+e^{i\alpha w}}{2\pi w
\kappa^{-1} (e^{\alpha\kappa}-1)}+\frac{1}{2\pi(i+w\kappa^{-1})}
\{e^{\alpha\kappa(1-iw\kappa)}\times \ \nonumber
\\ & &
F[1,1-iw\kappa^{-1},2-iw\kappa^{-1},e^{\alpha\kappa}]-e^{-\alpha\kappa(1-iw\kappa)}
\times \nonumber
\\ & & F[1,1-iw\kappa^{-1},2-iw\kappa^{-1},e^{-\alpha\kappa}]\}\ \ \
 \eea
Here,  even in the very low energy regime, the contribution of short
distances is not negligible. In fact, it is much bigger than the
thermal spectrum itself. To see this we can approximate
$I^B_{i\epsilon}(w\kappa^{-1}, \alpha\kappa)$ as \bea \label{aaa}
&&I^B_{i\epsilon}(w\kappa^{-1},\alpha\kappa)=\left( \frac{1}{\pi
\alpha\kappa}+\frac{\alpha\kappa}{12 \pi}-
\frac{(\alpha\kappa)^3}{720
\pi}+O((\alpha\kappa)^5)\right)\frac{\kappa}{w}-\frac{1}{2}\nonumber
\\&+&\left(\frac{\alpha\kappa}{2\pi}
-\frac{(\alpha\kappa)^3}{72\pi}+O((\alpha\kappa)^5)\right)\frac{w}{\kappa}-\left(\frac{(\alpha\kappa)^3}{72
\pi}+O((\alpha\kappa)^5)\right)(\frac{w}{\kappa})^3+...
 \ \ \ \ \eea
Note that in this case the dominant term is of order
$1/\alpha\kappa$, therefore \be
\lim_{\kappa^{-1}w\rightarrow0}\frac{I^B_{i\epsilon}(w\kappa^{-1},\alpha\kappa)}{(e^{2\pi
w\kappa^{-1}}-1)^{-1}}=\frac{2}{\alpha\kappa}\gg 1 \ . \ee A
similar behavior can be found for $w \approx w_{typical}$.

\begin{figure}[htbp]
\begin{center}
\includegraphics[angle=0,width=6.0in,clip]{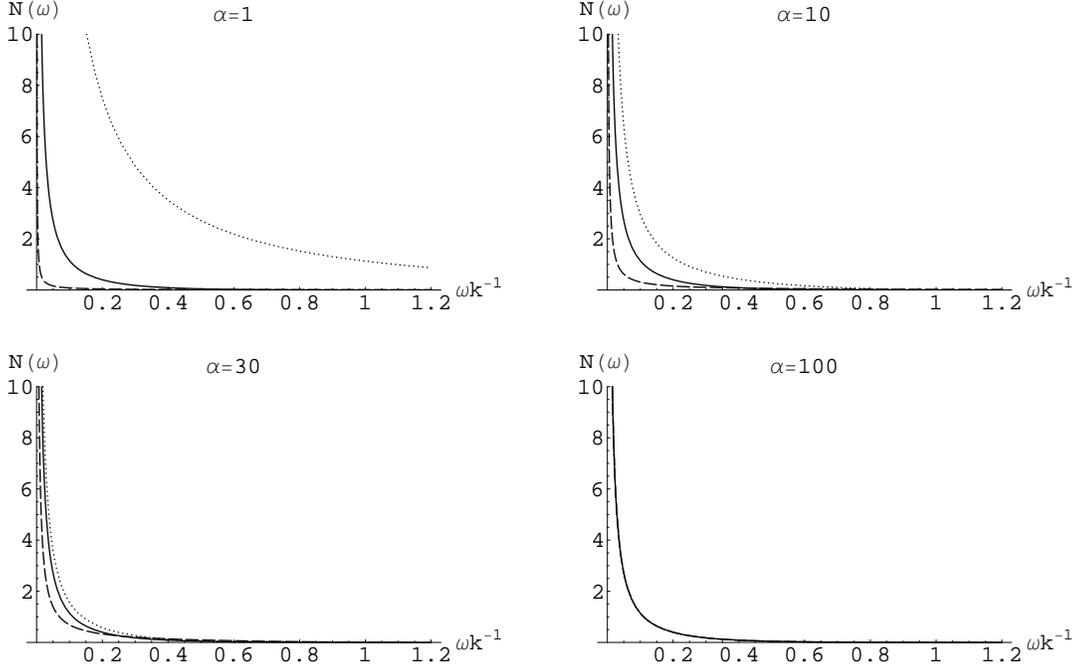}
\label{Fig2} \caption{Plot comparing the Planckian spectrum
$N(w,\kappa)=(e^{2\pi w \kappa^{-1}}-1)^{-1}$ (solid line) with the
contributions $I^B$ (dashed line) and $I^B_{i\epsilon}$ (dotted
line) coming from distances $|z|\lesssim \alpha$ according to the
normal-ordering prescription  and the $i\epsilon$ prescription,
respectively . We have taken $\kappa =0.1$ and $\alpha= 1, 10, 30$
and $100$ (in Planck units), respectively.}
\end{center}
\end{figure}

We illustrate the difference between both calculations in Fig.2.
With the normal-ordering prescription the short-distance
contribution is small, in contrast with the
$i\epsilon$-prescription.  We have chosen a large surface gravity
and different values of $\alpha$ to better show the effect in the
drawings. We clearly observe  that, although both prescriptions
lead to the thermal result when $\alpha \to +\infty$, they do the
job in very different ways.

\subsection{Fermions}

We shall extend the previous analysis to fermions. The integral involved is
\begin{eqnarray}
& & \ \ \ \ \ \ \ \ \ \ \ \ \ \ \ I^F(w\kappa^{-1},
\alpha\kappa)\equiv \frac{-i}{2\pi} \int_{-\alpha}^{+\alpha} dz
e^{-iwz} \left[\frac{(\frac{\kappa}{2})}{(\sinh
\frac{\kappa}{2}z)^2}-\frac{1}{z}\right]= \ \ \ \ \ \ \nonumber \\
&& \frac{Si(\alpha w)}{\pi}+\frac{1}{2 \pi(1+4 w^2\kappa^{-2})}
\{(-i+2 w\kappa^{-1}) (e^{-\alpha\kappa/2+i\alpha w}\times
\nonumber
\\ & &
F[1,\frac{1}{2}-iw\kappa^{-1},\frac{3}{2}-i w
\kappa^{-1},e^{-\alpha\kappa}] -
F[1,\frac{1}{2}-iw\kappa^{-1},\frac{3}{2}-i w
\kappa^{-1},e^{\alpha\kappa}]\times \nonumber \\
& &e^{\alpha\kappa/2-i\alpha w})+(i+2 w\kappa^{-1})
(e^{-\alpha\kappa/2-i\alpha w}
F[1,\frac{1}{2}+iw\kappa^{-1},\frac{3}{2}+i w
\kappa^{-1},e^{-\alpha\kappa}]\nonumber
\\ & &-e^{\alpha\kappa/2+i\alpha w}
F[1,\frac{1}{2}+iw\kappa^{-1},\frac{3}{2}+i w
\kappa^{-1},e^{\alpha\kappa}])\}\ \ \ . \eea See Fig. 3 for a
graphical representation. Taking into account that $\alpha\kappa
\ll1$ we can expand $I^F(w\kappa^{-1}, \alpha\kappa)$ as \be \label{eq:exp-fer}
I^F(w\kappa^{-1},\alpha\kappa)= \left(\frac{(\alpha\kappa)^3}{72 \pi
}+O[(\alpha\kappa)^5]\right)\frac{w}{\kappa}+O[(\alpha\kappa)^5](\frac{w}{\kappa})^3+...\ee
Note that the term proportional to $\kappa/w$, appearing in the
bosonic case, has disappeared. Therefore, for very low frequencies
\be \frac{I^F(w\kappa^{-1},\alpha\kappa)}{(e^{2\pi
w\kappa^{-1}}+1)^{-1}} \sim \frac{(\alpha\kappa)^3}{36 \pi
}\frac{w}{\kappa} \ll 1 \ . \ee This shows that the contribution of
ultrashort distances is negligible, like in the bosonic case.
Moreover, for typical Hawking frequencies, $w_{typical}= T_H$, we
have \be\frac{I^F(w_{typical}\kappa^{-1},\alpha\kappa)}{(e^{2\pi
w_{typical}\kappa^{-1}}+1)^{-1}} \sim 3\cdot 10^{-3}(\alpha\kappa)^3
 \ll 1\ . \ee This rate is again very small, but the
above expressions unravel the fact that the short-distance
contribution for fermions seems to be smaller than that of bosons.
For the latter the contribution of short distances is proportional
to the first power of $\kappa\alpha$ while for fermions it is the
third power.

\begin{figure}[htbp]
\begin{center}
\includegraphics[angle=0,width=6.0in,clip]{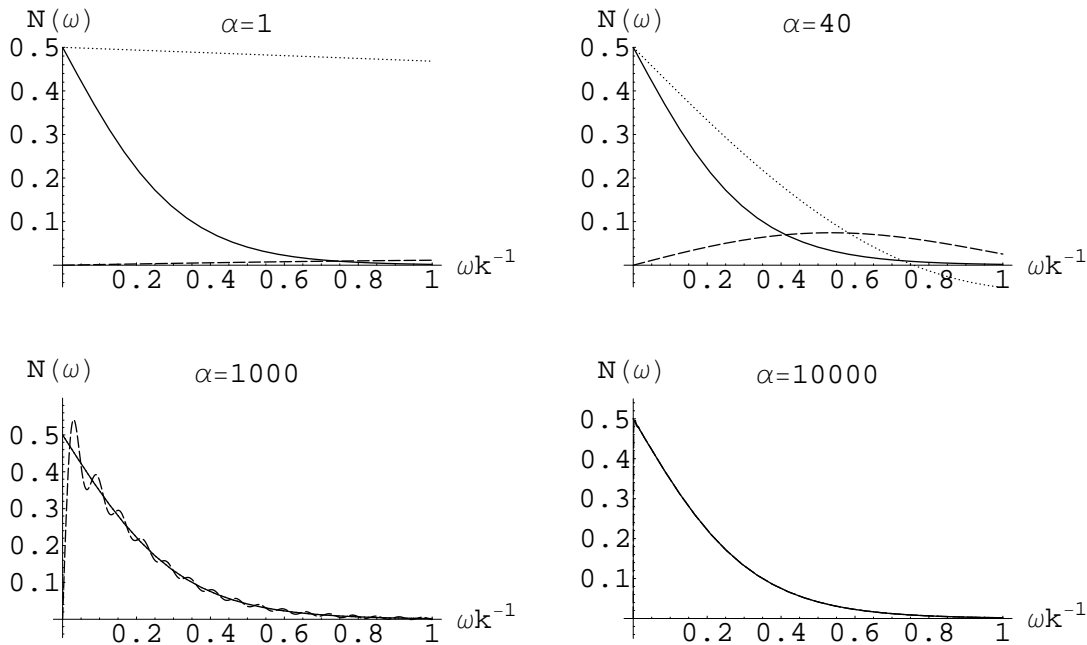}
\label{Fig3} \caption{Plot comparing the Dirac-Fermi distribution
(solid line) $N(w,\kappa)=(e^{2\pi w \kappa^{-1}}+1)^{-1}$ with the
contribution $I^F$ coming  from distances $|z|< \alpha$ according to
the normal-ordering prescription (dashed line). For completeness we
have also plotted the result obtained with the
$i\epsilon$-prescription (dotted line). We have taken $\kappa =0.1$
and $\alpha= 1, 40, 10^3$ and $10^4$ (in Planck units),
respectively.}
\end{center}
\end{figure}

Finally, let us give numerical estimates for  relevant
astrophysical black holes using the expansion (\ref{eq:exp-fer})
 \be\frac{I^F(w\kappa^{-1},\alpha\kappa)}{(e^{2\pi
w\kappa^{-1}}+1)^{-1}} \approx \frac{\alpha^3 w \kappa^2 (e^{2\pi
w\kappa^{-1}}+1)}{72\pi}\ . \ee For a black hole of three solar
masses, the relative contribution to the total Planckian spectrum
is, for $w=w_{typical}$, of order $10^{-118}\%$ and one must go to
frequencies of order $w/w_{typical}\approx 270$ to find
contributions $I^F(w\kappa^{-1}, l_P\kappa)$ of the same order as
the total spectrum. For primordial black holes, $M\sim 10^{15}$ g,
the relative contribution is  $10^{-62}\%$ at $w_{typical}$ and we
have to reach frequencies of order $142w_{typical}$ to get a
short-distance contribution of order of the thermal distribution.
In addition to the conclusions stressed in the bosonic case,
namely  the robustness of Hawking thermal radiation for
wavelengths of order of the size of the black hole, we have a new
result. Fermions seem to be less sensitive to ultrashort distance
physics than (spinless) bosons.

\section{Modifying the two-point functions at short-distances}

In the previous section, we have investigated the contribution to
the Hawking spectrum coming from distances $z<l_P$ at $I^+$ assuming that the
physical laws are not modified at such scales. We found that
potential deviations from thermality only manifest themselves at high
frequencies. One can legitimately wonder, however, why we looked
at distances at $I^+$ instead of at $I^-$,
where the sensitivity of the ``in'' state to short distances is
more apparent. In fact, imposing naively a cutoff at $I^-$ has
dramatic effects on the radiation due to the enormous redshift
caused by the horizon (see section \ref{sec:Bog}). We were motivated
to impose the cutoff at $I^+$ in order to find agreement with the view
offered by string theory. The purpose of this section is to shed light
on the roles played by distances at $I^+$ and $I^-$ by using a simple model with
a modified two-point function. We shall investigate the potential effects on
the radiation due to the modified short-distance behavior of the
matter field, supposedly coming from unknown physics at the Planck
scale. We shall see  how our model maintains the robustness of the Hawking
thermal spectrum at $I^+$, while at the same time being insensitive to sub-Planckian
distances at $I^-$.\\

Let us assume that the standard two-point function for the spin
zero ``in'' and ``out'' vacuum states at $I^-$ and $I^+$,
respectively, gets modified by new physics at very short distances
and becomes \bea
G^{in}|_{I^-}\equiv-\frac{1}{4\pi}\frac{1}{(v_1-v_2)^2} &\to&
G^{in}_{\alpha}|_{I^-}\equiv-\frac{1}{4\pi}\frac{1}{(v_1-v_2)^2 + \alpha^2} \nonumber \\
G^{out}|_{I^+}\equiv-\frac{1}{4\pi}\frac{1}{(u_1-u_2)^2} &\to&
G^{out}_{\alpha}|_{I^+}\equiv-\frac{1}{4\pi}\frac{1}{(u_1-u_2)^2 +
\alpha^2} \ , \eea where $\alpha$ is a parameter of order of the
Planck length: $\alpha \sim l_P$. With this modification the
expression (\ref{N-nordI-}) for the black hole particle production
becomes (we omit the transmission coefficients $t_l(w)$ and the
angular delta functions $\delta_{l_1 l_2}\delta_{m_1 m_2}$ since
they are also irrelevant for the discussion of this section)
 \bea \langle in|N_{i_1i_2}^{out}|in \rangle
&=&
4\int_{-\infty}^{v_H}
 dv_1 dv_2
u^{out}_{w_1}(v_1)u^{out*}_{w_2}(v_2)\times \nonumber
\\&&\left [-\frac{1}{4\pi}\frac{1}{(v_1-v_2)^2+ \alpha^2}-\frac{du}{dv}(v_1)\frac{du}{dv}(v_2)G^{out}_{\alpha}|_{I^-} \right ] \ , \eea
where $u^{out}_w$ and
$G^{out}_{\alpha}|_{I^-}$ are understood to be the ``out'' modes and the
``out'' two-point function, respectively, propagated
back to $I^-$. Since, according to the standard derivation, the propagation to $I^-$ implies a strong blueshift, the $u^{out}_w$ modes
and $G^{out}_{\alpha}$ might manifest some dependence on the
particular details of the modified theory, which are unknown to us. Thus, we see no simple way to
 estimate the form of the $u^{out}_w$ modes at $I^-$. For this reason, it is preferable to evaluate the particle production as an integral on $I^+$, as in (\ref{N-nordI+}),
\begin{eqnarray} \label{N12m}&&\langle in|N^{out}_{i_1i_2}|in \rangle =
 4\int_{I^+}du_1du_2  u^{out}_{w_1}(u_1)u^{out*}_{w_2}(u_2)\times \nonumber \\
&\times&  \left[ \frac{dv_1}{du_1}\frac{dv_2}{du_2}G^{in}_{\alpha}|_{I^+}
+\frac{1}{4\pi}\frac{1}{(u_1-u_2)^2+\alpha^2} \right] ,
\end{eqnarray}
where $G^{in}_{\alpha}|_{I^+}$ is understood to be the ``in'' two-point
function propagated to $I^+$. In this region we can use the
standard form of the ``out'' modes $u^{out}_{w_1}(u_1)=\frac{e^{-iw_1u_1}}{\sqrt{4\pi w}}$ since we are considering
emission frequencies much lower than the Planck frequency $w_P\sim 1/l_P$. We
still have to unravel the evolution of $G^{in}_{\alpha}$ to
evaluate the above expression. The modified short-distance physics
near the horizon could dramatically modify the evolution of the
two-point function, so that its form at $I^+$ could be rather
different from the standard one $G^{in}|_{I^-}$. However, we can make the
reasonable assumption that the propagation to $I^+$ is affected by new physics in such a way that
the short-distance behavior of $\frac{dv_1}{du_1}\frac{dv_2}{du_2}G^{in}_{\alpha}$ at $I^+$ is
identical to that of the two-point function for the ``out'' state
\be\label{Hcondition}
\lim_{u_1\to u_2}\frac{dv}{du}(u_1)\frac{dv}{du}(u_2)G^{in}_{\alpha}|_{I^+} \ \sim \ \lim_{u_1\to u_2} G^{out}_{\alpha}(u_1, u_2)|_{I^+} \ .
\ee
The above condition can be seen as a natural generalization of the
Hadamard condition, i.e., universality of the short-distance
behavior for all quantum states. The Hadamard condition, which
plays a pivotal role in  the algebraic formulation of QFT in
curved spacetime \cite{waldbook}, ensures the regularity of
expression (\ref{eq:N-nord}) to evaluate the Hawking radiation. \\
Let us see now how (\ref{Hcondition})  constraints the evolution of
${G}^{in}_{\alpha}$ from $I^-$ to ${I^+}$.  Note that
${G}^{in}_{\alpha}$ can be rewritten as \be \label{eq:G_aofG_in} {G}^{in}_{\alpha} =
\frac{G^{in}}{1+\alpha^2G^{in} } , \ee where $G^{in}$ is
the unmodified two-point function. Since, at late
times, $G^{in}$ evolves according to geometrical optics approximation 
\be \label{raytracing0}
\tilde{G}^{in}|_{I^+}\equiv\frac{dv}{du}(u_1)\frac{dv}{du}(u_2)G^{in}|_{I^+}=-\frac{1}{4\pi}\frac{\frac{dv_1}{du_1}\frac{dv_2}{du_2}}{(v(u_1)-v(u_2))^2}, 
\ee 
expression (\ref{eq:G_aofG_in}) suggests the following evolution for ${G}^{in}_{\alpha}$  
\be \label{Gin}
\tilde{G}^{in}_{\alpha}|_{I^+}\equiv\frac{dv}{du}(u_1)\frac{dv}{du}(u_2)G^{in}_{\alpha}|_{I^+}
=
-\frac{1}{4\pi}\frac{\frac{dv(u_1)}{du}\frac{dv(u_2)}{du}}{(v_1-v_2)^2
+ \alpha^2\frac{dv_1}{du_1}\frac{dv_2}{du_2}} \  . \ee This expression 
guarantees immediately the Hadamard condition (\ref{Hcondition}). We
should stress, however, that the evolution of the modified two-point
function itself is not equivalent, at least for very small point
separations $(u_2 -u_1)^2 \sim \alpha^2$, to the  ray tracing (or
geometrical optics approximation), which would produce instead
(\ref{raytracing}) (see later) and violate the Hadamard condition.
For larger separations $(u_2 -u_1)^2\gg\alpha^2$ the propagation agrees, as it must, with  standard
relativistic field theory and is driven by the large redshift
(implying then the usual geometrical optics approximation). Plugging this
expression in (\ref{N12m}) we obtain
\begin{eqnarray} \label{N12m2}&&\langle in|N^{out}_{i_1i_2}|in \rangle =
 \frac{-1}{4\pi^2\sqrt{\omega
_1\omega _2}}\int_{I^+}du_1du_2  e^{-i(w_1u_1-w_2u_2)} \nonumber \\
&\times&  \left[ \frac{\frac{dv(u_1)}{du}\frac{dv(u_2)}{du}}{
(v_1-v_2)^2 +\alpha^2\frac{dv(u_1)}{du}\frac{dv(u_2)}{du}}
-\frac{1}{(u_1-u_2)^2+\alpha^2} \right]\ .
\end{eqnarray}
It is worth noting that the modified term \be
-4\pi\tilde{G}^{in}_{\alpha}|_{I^+}\equiv\frac{dv_1}{du_1}\frac{dv_2}{du_2}\frac{1}{
(v_1-v_2)^2 +\alpha^2\frac{dv_1}{du_1}\frac{dv_2}{du_2}} \ , \ee
which can also be regarded as a transformation law under the
change $v=v(u)$, guarantees the absence of particle production
under the same group of symmetry transformations (Möbius
rescalings) as those of the theory with $\alpha =0$.\footnote{
Even more, it is what exactly guarantees the invariance of the
production rate, up to a shift on the emission frequency, under a
radial boost with rapidity $\xi$: $u \to \bar{u}=e^{\xi}u, v \to
\bar{v}=e^{-\xi}v$. }  Assuming that the geometry remains classical (the black hole scale $\kappa$ is well above the Planck scale $\alpha$), we can use in
(\ref{N12m2}) the expression $v(u)=v_H-\kappa^{-1}e^{-\kappa u}$,
which represents the relation between the ``in'' and ``out''
inertial coordinates. Performing then the integration in $u_2+u_1$,
we are left with $(z\equiv u_2 -u_1$)  \bea
\label{Norderingfinal0alpha} \langle in |N^{out}_{wlm}|in\rangle
&=& -\frac{1}{2\pi w} \int_{-\infty}^{+\infty} dz e^{-iwz}
\left[\frac{(\frac{\kappa}{2})^2}{(\sinh \frac{\kappa}{2}
z)^2+(\frac{\kappa}{2})^2\alpha^2}-\frac{1}{z^2+\alpha^2}\right ]
\ . \eea Finally, performing the integration in the complex plane,
the particle production rate becomes
\begin{eqnarray}\label{eq:Number-alpha}
\langle in|N^{out}_{wlm}|in \rangle=\frac{1}{(e^{2\pi
w\kappa^{-1}}-1)} \frac{1}{2w\alpha\sqrt{1-\alpha^2 \kappa^2/4}}
(e^{w\kappa^{-1}\theta} -e^{w\kappa^{-1}(2\pi
-\theta)})+\frac{e^{-w\alpha}}{2\alpha w}
\end{eqnarray}
where \be \theta = \arctan\frac{\alpha\kappa
\sqrt{1-\alpha^2\kappa^2/4}}{(1-\alpha^2\kappa^2/2)} \ . \ee The
thermal Planckian spectrum is smoothly recovered in the limit
$\alpha\to 0$. Moreover, for  $\alpha \sim l_P$, the deviation to
the thermal spectrum is negligible for small values of
$w\kappa^{-1}$. This deviation can be expanded as
\be
\frac{\langle
in|N^{out}_{wlm}|in \rangle}{(e^{2\pi w\kappa^{-1}}-1)^{-1}} \approx
1 - \frac{\alpha\kappa (e^{2\pi w\kappa^{-1}}-1)}{16 w\kappa^{-1}} \ .
\ee
 For astrophysical black holes, $\kappa \alpha\ll1$, the
second factor is negligible for frequencies up to $\sim 10^2w_{typical}$,
in complete agreement with  the results obtained in section \ref{sec:short}  (compare, for instance, with (\ref{eq:ratio3})).  \\

The above discussion shows that, despite the apparent
sensitivity of Hawking radiation to high energy physics (see section \ref{sec:Bog}), a
Planck-scale modification of the two-point function does not
necessarily imply a substantial change of the Planckian spectrum.
This is so, at least, if the modified two-point function obeys a modified
Hadamard-type condition. The simplest realization of this condition
turns out to be equivalent to the preservation of the powerful
conformal (Möbius) symmetry existing in the unmodified theory.
This seems an unavoidable requirement if the corrections to the Planckian
spectrum are to be in agreement with the results of string theory in the low-frequency limit $w\to 0$.
The effect of the generalized Hadamard condition is to constrain the short-distance behavior of
the propagated ``in'' two-point function, $\tilde{G}^{in}_{\alpha}|_{I^+}$, in such a way that it remains close to $-1/(4\pi \alpha^2)$ through its evolution to $I^+$, despite the large blueshift. In fact, $\tilde{G}^{in}_{\alpha}|_{I^+}$ is an observer-independent quantity in the limit $x_1\to x_2$, i.e., it tends to $-1/(4\pi \alpha^2)$ for any function $v=v(u)$. Note in passing that this condition is somewhat related to approaches to quantum gravity aimed at deforming Lorentz symmetry while keeping
the principle of relativity \cite{amelino-camelia-maguejo-smolin}.\\

To conclude, we  note that if the deformed two-point function at $I^-$
\be
G^{in}_{\alpha}|_{I^-}\equiv-\frac{1}{4\pi}\frac{1}{(v_1-v_2)^2 +
\alpha^2} \ ,
\ee
is naively  propagated (i.e., by ray tracing) to
$I^+$ as
\be \label{raytracing}
\tilde{G}^{in}_{\alpha}|_{I^+}=-\frac{1}{4\pi}\frac{\frac{dv_1}{du_1}\frac{dv_2}{du_2}}{(v(u_1)-v(u_2))^2
+ \alpha^2} \ , \ee where $\tilde{G}^{in}_{\alpha}|_{I^+}\equiv
\frac{dv}{du}(u_1)\frac{dv}{du}(u_2)G^{in}_{\alpha}|_{I^+}$, the
particle production rate is now time-dependent and the thermal
spectrum is lost for any  nonvanishing $\alpha$.

\section{Conclusions and final comments}

It is highly non-trivial \cite{jacobson9193} to truncate Hawking's
derivation of black hole radiance to account for unknown physics
at the Planck scale. A simple estimate of the contribution of virtual high
frequencies apparently shows that they are essential to
produce the thermal outcome. One can then change strategy and try
to evaluate the contribution of Planckian physics in position space,
which requires a rederivation of the Hawking calculation in
terms of two-point functions, as we have explicitly shown in
section \ref{sec:III}. When these two-point functions are treated in the
distributional sense, with the   usual $i\epsilon$ prescription,
one reproduces exactly the thermal result. However, one can
equivalently handle the divergence of the two-point function by
trivially taking normal ordering. The consistency of this procedure
is guaranteed by the Hadamard condition: the short-distance behavior
is universal for all physical states. The advantage of this second
option is that it offers a natural way to evaluate the contribution
of short distances at $I^+$ to Hawking radiation.\\

We have found that the contribution of short-distances at low
frequencies $w\ll \kappa$ is negligible. Our analysis allows us to
go further and investigate the short-distance contribution for
frequencies of order the Hawking temperature $T_H$ and beyond. We
find that the contribution of ultrashort distances is also
negligible for frequencies of order $T_H$. In fact,  for a black
hole of three solar masses we need to look at high frequencies,
$w/w_{typical}\approx 96$ (for bosons) or $w/w_{typical}\approx
270$ (for fermions), to find that the contribution of Planck
distances is of order of the total spectrum itself. This suggests
that Hawking thermal radiation is very robust, as it has been
confirmed in completely different analyses based on black hole
analogues;  in string theory (for large wavelength) in
near-extremal charged black holes; and  also in some models
of canonical quantum gravity \cite{kiefer}.  \\

One can legitimately ask why, in section IV, we evaluate distances
at $I^+$, instead of just at $I^-$, where the sensitivity of the
``in'' state to high energy scales is more apparent, as we showed
in section \ref{sec:Bog}. Our heuristic motivation is based on the
view offered by string theory, where the Hawking radiation is
obtained as the result of collisions between open string
excitations. In that approach, the standard large blueshift of
low-energy gravity theory does not seem to play the pivotal role
that it does in the pure semiclassical treatment. The fact that we
consider the fundamental Planck scale at $I^+$ does not immediately
guarantee that the Hawking radiation is kept unaltered from Planck-scale
physics. As we show in section IV, with the standard $i\epsilon$-prescription
the short-distance contribution to Hawking radiation is not negligible.
In contrast, with the normal-ordering prescription the bulk of the Hawking
effect is maintained at low frequencies, in agreement with the
results of string theory. \\

In addition to the above arguments we have approached the problem in
section V in a different way. We have considered an explicit
modification of the two-point function at the Planck scale.
Motivated by the crucial role played by the Hadamard condition in
the ordinary relativistic theory, we have assumed that the
short-distance behavior of the modified theory should also satisfy a
sort of generalized Hadamard condition ({\it universal} short-distance
behavior). The simplest realization of this idea turns out to be
equivalent to the preservation of the powerful conformal (Möbius)
symmetry existing in the unmodified theory. Armed with this condition,
the contribution to the particle production rate of the ``in'' and
``out'' two-point functions in (\ref{N12m2}) is similar when they are
compared in the same ultrashort range of distances, despite the
large blueshift horizon effect. As a result, the two contributions
compensate each other and lead to an emission spectrum very insensitive
to trans-Planckian physics.  The generalized Hadamard condition,
therefore, seems to be necessary to maintain the bulk of the Hawking
effect. Moreover, it is in this context that the apparent tension
between $I^+$ and $I^-$ to measure separations is elliminated since in both
we find the same finite short distance limit.\\

A last comment is now in order. In the string theory analysis one
has, at least, two relevant parameters: the surface gravity
$\kappa$ and the radius $r_g$ of the supersymmetric, charged black
hole. The surface gravity is assumed to be small, in comparison with
the inverse of the size of the black hole, i.e., $\kappa\ll 1/r_g$.
The emission frequency can reach $\kappa$, but can never reach $1/r_g$
(or become larger) to guarantee the validity of the string theory
calculation. Obviously the analysis of string theory excludes
astrophysical black holes of the Schwarzschild type (for which
$\kappa \sim 1/r_g$). Our results, however, suggest that one could
also expect string theory to predict in this case, in some subtle
way, agreement with Hawking's results for frequencies around
$1/r_g$ and, at least, a few orders beyond. This is so because we
do not observe any significant contribution to the thermal
spectrum coming from the short-distance region, where new physics
could arise, up to such high frequencies. This fact offers a very
non-trivial challenge for any quantum theory of gravity having
computational rules very different from those of semiclassical
gravity (as in string theory or background-independent approaches),
since when $w \sim 1/r_g$ the grey-body factors $\Gamma_i(w)$
cannot be computed analytically. They are only known numerically
\cite{page}, as a result of solving field wave equations in the
black hole background. String theory manages to account for the
greybody factors in the low-energy regime, where they admit an
analytic expression. In fact, for all spherically symmetric black
holes, the low-energy absorption cross section is proportional to
the area of the horizon \cite{das-gibbons-mathur, unruh}. But for
typical Hawking frequencies  the grey-body factors remain elusive
for any analytic treatment. Reobtaining them from such a different
computation would be extremely impressive.

\section*{Appendix A}

We will complete here the steps missing in the derivation that led
to the emission rate (\ref{eq:N-timedependent}). Using the wave
packets
 (\ref{eq:wavepacket}) we can express it as
\bea  & & \langle in |N^{out,\sigma}_{j_1n_1,j_2n_2}|in
\rangle=\int_0^{\Lambda}dw' \beta_{j_1n_1,w'} \beta_{j_2n_2,w'}^* =\\
\nonumber & & \frac{1}{\epsilon}
\int^{(j_1+1)\epsilon}_{j_1\epsilon} dw_1\
\int^{(j_2+1)\epsilon}_{j_2\epsilon} dw_2\ e^{2\pi iw_1
n_1/\epsilon}\ e^{-2\pi iw_2 n_2/\epsilon}\ \int_{0}^{\Lambda} dw'
\beta_{w_{1}w'} \beta_{w_{2}w'}^{*}\ . \eea Using
(\ref{integralw'2}) we get \bea  \langle in
|N^{out,\sigma}_{j_1n_1,j_2n_2}|in \rangle&=&\frac{1}{\epsilon}
\int^{(j_1+1)\epsilon}_{j_1\epsilon} dw_1\
\int^{(j_2+1)\epsilon}_{j_2\epsilon} dw_2\ e^{i\frac{2\pi w_1
n_1}{\epsilon}}\ e^{-i\frac{2\pi w_2 n_2}{\epsilon}} \nonumber
\\  & & t_l(w_1)t^*_l(w_2)\frac{e^{-i (w_1-w_2)v_H}}{2\pi\sqrt{w_1 w_2}} e^{-\pi\kappa^{-1}
\omega_1} i^{-i\kappa^{-1}(w_1-w_2)}\nonumber \\ & &
\Gamma(1+i\kappa^{-1}w_1) \Gamma(1-i\kappa^{-1}w_2)
\delta_\sigma[\kappa^{-1}(w_1-w_2)] \ . \ \   \eea This integral
can be estimated explicitly when the width $\epsilon$ of the
frequency interval $[j\epsilon, (j+1)\epsilon]$ is assumed, as
usual, small. In this case, the integral is essentially as follows
\begin{equation}\label{Nlambda1}
\langle in |N^{out,\sigma}_{j_1n_1,j_2n_2}|in \rangle \approx
\delta_{j_1j_2}
\frac{|t_l(w_j)|^2|\Gamma(1+i\kappa^{-1}w_j)|^2}{2\pi w_j}e^{-\pi
\kappa^{-1}w_j}e^{\frac{2\pi(n_1-n_2)w_j}{\epsilon}}I_{n_1n_2}(\sigma)
\end{equation}
where
\begin{equation}
I_{n_1n_2}(\sigma)=\frac{1}{\epsilon}\int_{-\epsilon/2}^{\epsilon/2}dx_1\int_{-\epsilon/2}^{\epsilon/2}dx_2
e^{i[\frac{2\pi n_1}{\epsilon}-v_H]x_1-i[\frac{2\pi
n_2}{\epsilon}-v_H]x_2-\pi
\kappa^{-1}(x_1+x_2)/2}\delta_\sigma[\kappa^{-1}(x_1-x_2)]
\end{equation}
and $x_{1,2}\equiv w_{1,2} -(j+1/2)\epsilon $. The factor
$\delta_{j_1j_2}$ in (\ref{Nlambda1}) is due to the role of
$\delta_\sigma$, which selects frequencies on a very narrow band
of order $|w_1-w_2|\sim\kappa\sigma$. For this reason, it is also
convenient to introduce a new variable $y=x_1-x_2$ and rewrite
$I_{n_1n_2}(\sigma)$ as follows
\begin{equation}
I_{n_1n_2}(\sigma)\approx
\frac{1}{\epsilon}\int_{-\epsilon/2}^{\epsilon/2}dx_1
e^{\frac{2\pi(n_1-n_2)x_1}{\epsilon}}
\int_{x_1-\epsilon/2}^{x_1+\epsilon/2}dy e^{i[\frac{2\pi
n_2}{\epsilon}-v_H]y}\delta_\sigma[\kappa^{-1}y]
 \ . \end{equation}
In writing this we have neglected the term
$e^{\pi\kappa^{-1}(x_1+x_2)/2}$ which is almost constant (unity)
over the integral. We can now estimate the integral over $y$
having in mind that $\delta_\sigma$ is very well approximated by a
square step of width $\pi\kappa\sigma$ and height $1/(\pi\sigma)$
centered at $y=0$. This means that the main contribution comes
from the interval
$[-\frac{\pi\kappa\sigma}{2},\frac{\pi\kappa\sigma}{2}]$. This
fact makes the outcome of the integral independent of $x_1$, which
also allows us to perform the integral in $x_1$. Putting all
together we find
\begin{equation}
I_{n_1n_2}(\sigma)\approx
\kappa\delta_{n_1n_2}\frac{\sin\left[\left(\frac{2\pi
n_2}{\epsilon}-v_H\right)\frac{\pi\kappa\sigma}{2}\right]}{\left[\left(\frac{2\pi
n_2}{\epsilon}-v_H\right)\frac{\pi\kappa\sigma}{2}\right]}
\end{equation}
Plugging this result back into (\ref{Nlambda1}) we find (\ref{eq:N-timedependent}). \\

We will now briefly consider the effect of introducing the cutoff
in frequencies in a different way. The cutoff was introduced in
(\ref{integralw'2}) in the form
\begin{equation}
\int_{-\infty}^\infty d\log[w/\kappa]
e^{-i\kappa^{-1}(w_1-w_2)\log[w/\kappa]}\to
\int_{-\log[\Lambda/\kappa]}^{\log[\Lambda/\kappa]}
d\log[w/\kappa] e^{-i\kappa^{-1}(w_1-w_2)\log[w/\kappa]}
\end{equation}
We will now consider the change
\begin{equation}
\int_{-\log[\Lambda/\kappa]}^{\log[\Lambda/\kappa]} d\lambda
e^{-i\kappa^{-1}(w_1-w_2)\lambda}\to \int_{-\infty}^{\infty} d\lambda
e^{-i\kappa^{-1}(w_1-w_2)\lambda}e^{-( \lambda/\tilde{\Lambda})^2}
\end{equation}
where $\tilde{\Lambda}$ must be of order
$\sim\log[\Lambda/\kappa]$. This modification leads to a
redefinition of $\delta_\sigma$
\begin{equation}\label{eq:tildesigma}
\delta_{\tilde{\sigma}}[\kappa^{-1}(w_1-w_2)]=\frac{\exp\left(-\left[\frac{\kappa^{-1}(w_1-w_2)}{2\tilde{\sigma}}\right]^2\right)}{2\tilde{\sigma}\sqrt{\pi}}
\end{equation}
which in the limit $2\tilde{\sigma}\to0$ also becomes Dirac's
delta function. One can then proceed as above and define the
corresponding function $I_{n_1n_2}(\tilde{\sigma})$, which this
time can be evaluated extending up to infinity the limits of
integration over the variable $y=x_1-x_2$. This leads to
\begin{equation}
I_{n_1n_2}(\tilde{\sigma})=
\kappa\delta_{n_1n_2}e^{-\left[\left(\frac{2\pi
n_2}{\epsilon}-v_H\right)\kappa\tilde{\sigma}\right]^2}
\end{equation}
The corresponding emission rate is now
\begin{equation}
\langle in
|N^{out,\sigma}_{j_1n_1,j_2n_2}|in\rangle=\delta_{j_1j_2}\delta_{n_1n_2}
\frac{|t_l(w_j)|^2}{e^{2\pi \kappa^{-1} w_j} -
1}e^{-\left[\left(\frac{2\pi
n_2}{\epsilon}-v_H\right)\kappa\tilde{\sigma}\right]^2}
\end{equation}
This expression is always positive definite and exhibits the same
decay rate as (\ref{eq:N-timedependent}) if we identify
$2\tilde{\sigma}$ with $\sigma$, which in fact is the right choice
for the definition of (\ref{eq:tildesigma}).

\section*{Appendix B}

We will proceed now to solve the massless Dirac equation in a
curved background with spherical symmetry. The equation to solve
is\footnote{For  earlier references see \cite{brill-wheeler}, and
for a more advanced treatment (no needed for the purposed of this
paper) see \cite{chandrasekar}.}
\begin{equation}\label{eq:Dirac=0}
\gamma^{\mu}\nabla_\mu \psi=0
\end{equation}
where $\gamma^\mu=\gamma^a V_a^\mu(x)$ satisfy
$\{\gamma^\mu,\gamma^\nu\}=2g^{\mu\nu}$,
$\{\gamma^a,\gamma^b\}=2\eta^{ab}$ and $V_a^\mu
V_b^\nu\eta^{ab}=g^{\mu\nu}$ represent the {\it
vierbeins}\footnote{Due to our convention for the metric signature
the $\gamma^a$ matrices should verify the conditions
$(\gamma^0)^2=-I$,  $(\gamma^i)^2=I$. However, to agree with the
standard notation for Dirac matrices in this appendix we have
flipped  the metric signature to $(+,-,-,-)$. This is, however,
irrelevant for the  computations  carried out in the body of this
paper.}. Note that $\nabla_\mu \psi=(\partial_\mu-\Gamma_\mu)\psi$
where $\Gamma_\mu=-\frac{1}{4}\gamma^b\gamma^cV_b^\nu\nabla_\mu
V_{\nu c}$ represents the spin connection. We will take the curved
space line element $ds^2=e^{2\rho}dx^+dx^--r^2d\Omega^2$, with
$dx^+dx^-=dt^2-dr^{*2}$, and $\eta_{ab}=diag(1,-1,-1,-1)$.
Introducing the ansatz $\psi=\frac{e^{-\rho/2}}{r}\Phi$ and making
the simplest choice for  vierbeins (i.e., to be parallel to the
unit vectors in $t, r^*, \theta, \phi$ directions), the Dirac
equation (\ref{eq:Dirac=0}) becomes
\begin{equation}\label{eq:Dirac=1}
\gamma^aV_a^i\partial_i\Phi+\frac{1}{r}\left[\frac{\gamma^2}{\sin^{1/2}\theta}\partial_\theta\sin^{1/2}\theta+
\frac{\gamma^3}{\sin\theta}\partial_\phi\right]\Phi=0
\end{equation}
where the index $i$ runs over the non-angular variables. Since for $x^\pm=t\pm r^*$ we have
$\gamma^aV_a^i\partial_i=e^{-\rho}[\gamma^0\partial_t+\gamma^1\partial_{r^*}]$, (\ref{eq:Dirac=1}) can be written in the more familiar form
\begin{equation}\label{eq:Dirac=2}
\partial_t\Phi=-\gamma^0\gamma^1\left[\partial_{r^*}+\frac{e^\rho}{r}\left(\frac{\gamma^2\gamma^1}{\sin^{1/2}\theta}
\partial_\theta\sin^{1/2}\theta+\frac{\gamma^3\gamma^1}{\sin\theta}\partial_\phi\right)\right]\Phi
\end{equation}
The angular part of this equation can be reexpressed as
$e^{\rho}\gamma^0K/r$:
\begin{equation}\label{eq:Dirac=3}
\partial_t\Phi=-\gamma^0\gamma^1\left[\partial_{r^*}-\frac{e^\rho}{r}\gamma^0 K\right]\Phi
\end{equation}
where the operator $K$ \be
K=\gamma^0\left(\frac{\gamma^1\gamma^2}{\sin^{1/2}\theta}
\partial_\theta\sin^{1/2}\theta+\frac{\gamma^1\gamma^3}{\sin\theta}\partial_\phi\right)
\ee commutes with the Dirac equation as well as  $\vec{J}^2$ and
$J_3$ and, therefore, its eigenvalues can be used to characterize
the angular part $\chi_{m_j\kappa_j}$ of the modes:
$K\chi_{m_j\kappa_j}=(-\kappa_j)\chi_{m_j\kappa_j}$, with
$\kappa_j^2=(j+\frac{1}{2})^2$. Moreover the eigenfunctions
$\chi_{m_j\kappa_j}$ admit the following decomposition
$\chi_{m_j\kappa_j}=c^+\chi^+_{m_j\kappa_j}+c^-\chi^-_{m_j\kappa_j}$,
with
\begin{eqnarray}\label{eq:spinorangular}
\chi^+_{m_j\kappa_j}&=&\left[\begin{array}{c}  \eta(\hat{r})^{m_j}_{\kappa_j}\\
 \\  0\end{array}\right]\\
\chi^-_{m_j\kappa_j}&=&\left[\begin{array}{c}  0\\
 \\  \eta(\hat{r})^{m_j}_{-\kappa_j}\end{array}\right]
\end{eqnarray}
Therefore, in a stationary spacetime,
$\rho=\rho(r)$, a general solution can then be expressed as
\begin{eqnarray}\label{eq:spinor}
\psi_{w \kappa_j m_j}(x)=\frac{e^{-\rho/2}e^{-
iwt}}{r}\left[\begin{array}{c}G_{w\kappa_j}(r)
\eta(\hat{r})^{m_j}_{\kappa_j}
\\
 \\ -iF_{w\kappa_j}(r)\sigma^1\eta(\hat{r})^{m_j}_{\kappa_j} \end{array}\right]
\end{eqnarray}
where we have used that $\sigma^1\eta(\hat{r})^{m_j}_{\kappa_j}=\eta(\hat{r})^{m_j}_{-\kappa_j}$ and the functions  $F_{w\kappa_j}(r^*)$ and $G_{w\kappa_j}(r^*)$
satisfy the following equations (see also \cite{unruh})
\begin{eqnarray}\label{eq:radialf}
\partial_{r^*} G_{w\kappa_j}&=&-\frac{e^\rho}{r}\kappa_jG_{w\kappa_j}+w F_{w\kappa_j}\\
\partial_{r^*} F_{w\kappa_j}&=&\frac{e^\rho}{r}\kappa_jF_{w\kappa_j}-w G_{w\kappa_j}\label{eq:radialg}
\end{eqnarray}
Adding the time-dependent part, the above equations lead to plane-wave
solutions $\sim e^{- iw(t\pm r^*)}=e^{- iwx^\pm}$ for all $\kappa_j$ as $r\to \infty$.\\

We note that the form of the eigenfunctions $\chi_{m_j\kappa_j}$
can be worked out immediately if the vierbeins are chosen to be
parallel to unit vectors in the standard $t, x, y, z$
directions\footnote{With this orientation for the vierbeins $K$
can be written as $K=\gamma^0(I+2\vec{S}\cdot\vec{L})$, which is
the standard form of this operator in Minkowski space.}. The
bispinors $\eta_{m_j\kappa_j}$ can be constructed, as it is
well-known, using the Clebsch-Gordon rules for addition of angular
momentum in terms of spherical harmonics and two-component
spinors, and the result is
\begin{eqnarray}
\eta(\hat{r})^{m_j}_{\kappa_j<0}=\left[\begin{array}{c}
                              \sqrt{\frac{j+m_j}{2j}}\ Y_{j-1/2}^{m_j-1/2}(\theta,\phi) \\
                               \sqrt{\frac{j-m_j}{2j}}\ Y_{j-1/2}^{m_j+1/2}(\theta,\phi)
                                \end{array}\right]
\end{eqnarray}
and
\begin{eqnarray}
\eta(\hat{r})^{m_j}_{\kappa_j>0}=\left[\begin{array}{c}
                              \sqrt{\frac{j+1-m_j}{2j+2}}\ Y_{j+1/2}^{m_j-1/2}(\theta,\phi) \\
                               -\sqrt{\frac{j+1+m_j}{2j+2}}\ Y_{j+1/2}^{m_j+1/2}(\theta,\phi)
                                \end{array}\right]
\end{eqnarray} \\

With the modes given in (\ref{eq:spinor}) conveniently normalized, the quantized Dirac field can be expanded in modes as
\begin{equation}
\psi(x)=\sum_{\kappa_j m_j}\int dw \left[a_{w \kappa_j m_j}u_{w
\kappa_j m_j}(x)+b^\dagger_{w \kappa_j m_j}v_{w \kappa_j
m_j}(x)\right]
\end{equation}
where $u_{w \kappa_j m_j}(x)$ and $v_{w \kappa_j m_j}(x)$
represent positive and negative energy solutions respectively.
On the other hand, since we are dealing with massless spinors, it
is necessary, on physical grounds, to use states with well defined
helicity. In particular, left-handed spinors can be obtained from
(\ref{eq:spinor}) by projecting with
$P_L=\frac{1}{{2}}(I-\gamma^5)$, where
$\gamma^5=i\gamma^0\gamma^1\gamma^2\gamma^3$. We will therefore be
working with the (normalized) modes $\psi^L_{w j
m_j}=\frac{1}{\sqrt{2}}(\psi_{w |\kappa_j| m_j}- \psi_{w
-|\kappa_j| m_j})$.\\

We will now carry out the calculations that lead to
(\ref{eq:N-fermions}) (adapted now for chiral spinors). First
thing to note is that the propagated backwards mode
(\ref{eq:OUT-mode}) contains a term of the form $\sqrt{du(v)/dv}$.
A simple way to realize why this term arises is that it is
necessary to ensure the invariance of the scalar product under
time evolution. Putting aside backscattering effects, the Dirac
scalar product for out modes can be written, equivalently, as \be
\int_{I^+} d\Omega dur^2\bar{u}^{out}\gamma_+u^{out} = \int_{I^-}
d\Omega r^2dv\bar{u}^{out}\gamma_-u^{out}\ . \ee The above
equality requires, up to relative signs in the spinor components,
that \be u^{out}(v)|_{I^+}=
\sqrt{du(v)/dv}\Theta(v_H-v)u^{out}(u)|_{I^-} \ . \ee Note that
the factor $e^{-\rho/2}$ in (\ref{eq:spinor}) also signals this
behavior. Since the spinor $\psi(x)$ does behave as a scalar under
general changes of coordinates, it follows that the functions $F$
and $G$ must somehow compensate the change in $e^{-\rho/2}$ under
conformal transformations.

 Let us now focus on the integration over the angular
variables prior to (\ref{eq:N-fermions}). This integration can be
readily performed if we put the result of (\ref{eq:2Pspinor}) into
(\ref{eq:ai+ajspinor}). We then find
\begin{eqnarray}\label{eq:proof1}
\langle in|N_{i_1 i_2}|in\rangle &=&\sum_k\int_{I^-}
dv_2r_2^2d\Omega_2
\left(\bar{u}^{out,L}_{i_2}(x_2)\frac{[\gamma^0-\gamma^1]}{2}v^{in,L}_k(x_2)\right)\times\nonumber
\\ & &\times\int_{I^-}  dv_1
r_1^2d\Omega_1\left(\bar{v}^{in,L}_{k}(x_1)\frac{[\gamma^0-\gamma^1]}{2}u^{out,L}_{i_1}(x_1)\right)
\end{eqnarray}
where the indices $i_1,i_2$ and $k$ denote $(w,j,m_{j})$. Using
the modes of eqs.(\ref{eq:OUT-mode}) and (\ref{eq:IN-mode}) it is
immediate to verify that
\begin{eqnarray}\label{eq:proof2}
\int d\Omega_2\bar{u}^{out,L}_{i_2}(x_2)\frac{[\gamma^0-\gamma^1]}{2}v^{in,L}_k(x_2)&=&
\frac{t_{j_2}^*(w_{2})}{2\pi r_2^2}\sqrt{\frac{du(v)}{dv}} \Theta(v_H-v)\times \nonumber\\
&\times& e^{iw_2u(v_2)+iwv_2}\delta_{m_{j_2}m_k}\delta_{j_2 j_{k}}
\end{eqnarray}
where we have used that $\int d\Omega \
\eta^{m_j\dagger}_{\kappa_j}(\hat{r})\eta^{m_{j'}}_{\kappa_{j'}}(\hat{r})=
\delta_{m_{j}m_{j'}}\delta_{\kappa_j\kappa_{j'}}$.  An analogous
calculation applies to the second factor in (\ref{eq:proof1}).
Plugging these results back into (\ref{eq:proof1}) we obtain
\begin{eqnarray}
\langle in|N_{i_1i_2}|in\rangle &=&
\delta_{m_{j_1}m_{j_2}}\delta_{{j_1}{j_2}}\frac{t_{j_1}(w_{1})t_{j_2}^*(w_{2})}{4\pi^2}\int_{-\infty}^{v_H}dv_1dv_2
\\& &\sqrt{\frac{du(v_1)}{dv}\frac{du(v_2)}{dv}}
e^{-iw_1u(v_1)+iw_2u(v_2)}\int_0^\infty dw e^{-iw(v_1-v_2)}
\nonumber
\end{eqnarray}
There remains to perform the integration in $w$, which yields
\begin{equation}
\int_0^\infty dw e^{-iw(v_1-v_2)}= \lim_{\epsilon\to
0}\frac{-i}{(v_1-v_2-i\epsilon)}
\end{equation}
and leads to the sought-after result. \\

\section*{Acknowledgements}
I.Agulló thanks MEC for a FPU fellowship and R.M. Wald for his
kind hospitality at the University of Chicago. This work has been
partially supported by grants FIS2005-05736-C03-03 and EU network
MRTN-CT-2004-005104. G.J.Olmo and L.Parker have been supported by
NSF grants PHY-0071044 and PHY-0503366. I.Agulló also thanks  R.M.
Wald  for many discussions and useful comments. J. Navarro-Salas
thanks P. Anderson, R. Balbinot, A. Fabbri and R. Parentani for
interesting discussions.


\begin{thebibliography}{99}

\bibitem{hawk1}
S.W. Hawking, Nature 248 (1974) 30 ;
S. W. Hawking, {\it Comm. Math. Phys.} {\bf 43}199 (1975);
S. W. Hawking, {\it Phys. Rev.} {\bf D14}, 2460 (1976)

\bibitem{parkerwald75}
L. Parker  {\it Phys. Rev.} D {\bf 12} 1519 (1975);
R. M. Wald {\it Commun. Math. Phys.} {\bf 45} 9 (1975)

\bibitem{bardeen}
J.M. Bardeen, B. Carter and S.W. Hawking, {\it Commun. Math. Pys.} {\bf 31}, 181 (1973)

\bibitem{bekenstein}
J.D. Bekenstein, {\it Phys. Rev. D} {\bf 7}, 2333 (1973);
{\bf 9}, 3292 (1974)

\bibitem{waldbook}
R. M. Wald,  {\it Quantum field theory in curved spacetime and
black hole thermodynamics}, CUP, Chicago (1994); {\it Living
Rev.Rel.} {\bf 4}, 6 (2001)

\bibitem{frolov-novikov}
V.P. Frolov and I.D. Novikov, {\it Black hole physics}, Kluwer Academic Publishers, Dordrecht (1998)

\bibitem{icp2005}
A. Fabbri  and J.Navarro-Salas,  {\it Modeling black hole evaporation}, ICP-World
Scientific, London (2005)

\bibitem{jacobson9193}
T. Jacobson, {\it Phys. Rev.} D {\bf 44}
1731 (1991); {\it Phys. Rev} D {\bf 48} 728 (1993)

\bibitem{fredenhagen-haag90}
K. Fredenhagen and R. Haag, {\it Commun. Math. Phys.} {\bf 127} 273 (1990)

\bibitem{strings}
C. Callan and J.  Maldacena, {\it Nucl.Phys.} {\bf B475}, 645 (1996);
A. Dhar, G. Mandal and S. R. Wadia,{\it Phys. Lett.} {\bf B388}, 51 (1996);
S. Das and S. Mathur,{\it Nucl. Phys.} {\bf B478}, 561 (1996)

\bibitem{reviews}
J. M. Maldacena, {\it Nucl. Phys. Pro. Supp.}, {\bf 61A}, 111 (1998), hep-th/9705078;
A. W. Peet, {\it TASI lectures on black holes in string theory}, hep-th/0008241;
J. R. David, G. Mandal and S. R. Wadia, {\it Phys. Rep.} {\bf 369} 549 (2002)


\bibitem{maldacena-strominger}
J. Maldacena and A. Strominger, {\it Phys. Rev.} {\bf D56}, 4975 (1997)

\bibitem{parker-toms}
L. Parker and D. J. Toms,{\it Principles and applications of quantum field theory in curved
spacetime}, Cambridge University Press (to be published)

\bibitem{unruh95}
W.G. Unruh, {\it Phys. Rev.} D {\bf 51} 2827 (1995)

\bibitem{bmps-cj}
R. Brout, S. Massar, R. Parentani and P. Spindel,{\it Phys. Rev.} D {\bf 52} 4559 (1995);
S. Corley and T. Jacobson, {\it Phys. Rev.} D {\bf 54} 1568 (1996);
{\it Phys.Rev.} D {\bf 59} 124011 (1999);
S. Corley, {\it Phys. Rev.} D {\bf 57} 6280 (1998);
R. Balbinot, A. Fabbri, S. Fagnocchi and R. Parentani, Riv. Nuovo Cimento {\bf 28}, 1 (2005),gr-qc/0601079

\bibitem{Parker77}
L.Parker, in ``Asymptotic structure of space-time'', ed. by F.P.Esposito and L.Witten, Plenum Press,N.Y.(1977), see page 195

\bibitem{Kay-Wald}
B.S. Kay and R.M. Wald, {\it Phys. Rep.} {\bf 207}, 59 (1991)

\bibitem{agullo-navarro-salas-olmo06}
I. Agullo, J. Navarro-Salas and G.J. Olmo, {\it Phys. Rev. Lett.}  {\bf 97}, 041302 (2006)

\bibitem{amelino-camelia-maguejo-smolin} G. Amelino-Camelia, {\it Int. J.
Mod. Phys. D} {\bf 11} 35 (2002). J. Maguejo and L. Smolin, {\it
Phys. Rev. Lett.} {\bf 88}, 190403 (2002)

\bibitem{kiefer}  C. Kiefer, J. Mueller-Hill, T.P. Singh and C. Vaz, {\it Hawking radiation from the
 quantum Lemaitre-Tolman-Bondi
model}, gr-qc/0703008

\bibitem{page}
D.N. Page, {\it Phys. Rev.} D {\bf 13}, 198 (1976)

\bibitem{das-gibbons-mathur}
S. Das, G. Gibbons and S. Mathur,{\it Phys. Rev. Lett.} {\bf 78}, 417 (1997)



\bibitem{unruh}
W.G. Unruh, {\it Phys. Rev.} D {\bf 14}, 3251 (1976)

\bibitem{brill-wheeler}
D.R. Brill and J.A. Wheeler, {\it Rev. Mod. Phys.} {\bf 29}, 465 (1957);
D.G. Boulware, {\it Phys. Rev. } D {\bf 12}, 350 (1975)

\bibitem{chandrasekar}
S. Chandrasekar, {\it The mathematical theory of black holes}, Oxford University Press, New-York (1983)




\end{thebibliography}
\end{document}